\documentclass[%
 reprint, 
preprint,
superscriptaddress,
 amsmath,amssymb,
 aps,
onecolumn,
]{revtex4-2}
\usepackage{pgffor}
\usepackage[utf8]{inputenc}
\usepackage{graphicx}
\usepackage{dcolumn}
\usepackage{bm}
\usepackage{hyperref}
\usepackage{color}
\usepackage[justification=justified,
   format=plain]{caption}
\usepackage{subcaption}
\usepackage{amsmath}
\usepackage{amssymb}
\usepackage{amsthm}
\usepackage{csquotes}
\usepackage{amsfonts}
\usepackage{physics}
\usepackage[english]{babel}
\usepackage[version=3]{mhchem}

\begin{document}

\title{Interlayer exciton polaritons in homobilayers of transition metal dichalcogenides}
\author{Jonas K. König}
\email{jonas.koenig@physik.uni-marburg.de}
\affiliation{Fachbereich Physik, Philipps-Universit\"at, Marburg, 35032, Germany}
\author{Jamie M. Fitzgerald}
\affiliation{Fachbereich Physik, Philipps-Universit\"at, Marburg, 35032, Germany}
\affiliation{Department of Physics, Chalmers University of Technology, SE-412 96 Gothenburg, Sweden}
\author{Joakim Hagel}
\affiliation{Department of Physics, Chalmers University of Technology, SE-412 96 Gothenburg, Sweden}
\author{Daniel Erkensten}
\affiliation{Department of Physics, Chalmers University of Technology, SE-412 96 Gothenburg, Sweden}
\author{Ermin Malic}
\affiliation{Fachbereich Physik, Philipps-Universit\"at, Marburg, 35032, Germany}
\affiliation{Department of Physics, Chalmers University of Technology, SE-412 96 Gothenburg, Sweden}
\begin{abstract} Transition metal dichalcogenides integrated within a high-quality microcavity support well-defined exciton polaritons. While the role of intralayer excitons in 2D polaritonics is well studied, interlayer excitons have been largely ignored due to their weak oscillator strength. Using a microscopic and material-realistic Wannier-Hopfield model, we demonstrate that MoS$_2$ homobilayers in a Fabry-Perot cavity support polaritons that exhibit a large interlayer exciton contribution, while remaining visible in linear optical spectra. Interestingly, with suitable tuning of the cavity length, the hybridization between intra- and interlayer excitons can be 'unmixed' due to the interaction with photons. We predict formation of polaritons where $>90$\% of the total excitonic contribution is stemming from the interlayer exciton. Furthermore, we explore the conditions on the tunneling strength and exciton energy landscape to push this to even $100$\%. Despite the extremely weak oscillator strength of the underlying interlayer exciton, optical energy can be effectively fed into the polaritons once the critical coupling condition of balanced radiative and scattering decay channels is met. These findings have a wide relevance for fields ranging from nonlinear optoelectronic devices to Bose-Einstein condensation.\end{abstract}

\maketitle
\newpage
\section{Introduction}
Van der Waals heterostructures consisting of vertically stacked transition-metal dichalcogenides (TMDs) are characterized by bright intra- \cite{chernikov2014exciton,perea2022exciton} and almost dark interlayer excitons \cite{rivera2015observation, calman2018indirect, jiang2021interlayer}. The former corresponds to a configuration where the constituent electron and hole are both in the same TMD layer, and the latter to where the electron and hole are spatially separated in different layers. Thanks to binding energies of more than $100$ meV \cite{chernikov2014exciton,ovesen2019interlayer,merkl2019ultrafast,lorchat2021excitons}, both species of excitons are stable at room temperature, offering potential for applications in combined excitonic and nanophotonic devices \cite{krasnok2018nanophotonics}. Intralayer excitons possess a large oscillator strength, which enables them to strongly couple with light when integrated within a microcavity \cite{schneider2018two,perea2022exciton}. This leads to hybrid exciton-photon quasi-particles, known as polaritons. They are characterized by a Rabi splitting in linear optical spectra \cite{kavokin2003thin}.  The existence of these quasi-particles has been experimentally established in monolayers \cite{Dufferwiel2015,liu2016strong} and theoretically predicted in type-II heterostructures \cite{latini2019cavity, Fitzgerald2022}. In contrast, pure interlayer excitons have an oscillator strength multiple orders of magnitude smaller \cite{ross2017interlayer} and do not enter the strong coupling regime, although a modification of the emission rate has been demonstrated via the Purcell effect \cite{forg2019cavity}. Because of their spatially indirect nature, interlayer exciton possess long population recombination lifetimes up to hundreds of nanoseconds \cite{rivera2015observation,calman2018indirect,forg2019cavity,miller2017long} as well as a permanent out-of-plane dipole moment \cite{rivera2015observation}. The latter leads to increased nonlinearities via repulsive dipole-dipole interactions \cite{nagler2017interlayer,erkensten2021exciton}, and offers the promise of an electrical control of exciton propagation/trapping \cite{unuchek2018room,jauregui2019electrical} and energy/optics via the Stark effect \cite{leisgang2020giant,lorchat2021excitons,peimyoo2021electrical,hagel2022electrical}.

In contrast to type-II heterobilayers, naturally stacked homobilayers exhibit a significant tunneling of holes between the TMD layers at the K points of the Brillouin zone \cite{fang2014strong,alexeev2019resonantly}. This significantly modifies the bilayer eigenstates and leads to hybridized excitons which are partially inter- and intralayer-like \cite{brem2020hybridized}, with an apportioned oscillator strength \cite{gerber2019interlayer}. Crucially, this mixing leads to bright hybrid excitons with a significant interlayer character, which have been observed, for example, in $H_h^h$-stacked \ce{MoS2} homobilayers \cite{leisgang2020giant,lorchat2021excitons,peimyoo2021electrical}. It is therefore highly interesting to inquire how homobilayers behave when incorporated within a high-quality optical microcavity. In particular, can exciton polaritons inherit this interlayer component? Such ``dipolaritons", i.e., a three-way superposition of both direct and indirect excitons with photons, have been studied in the context of electrically-biased asymmetric coupled quantum wells in microcavities \cite{cristofolini2012coupling} and single quantum wells in photonic waveguides \cite{rosenberg2018strongly}, in both cases constructed from traditional semiconductors. TMD homobilayers offer a potentially simpler system that do not need an external electric field and support interlayer exciton polaritons over a wider range of temperatures due to their larger exciton binding energies. Two recent experimental studies of the naturally-occurring $H_h^h$-stacked \ce{MoS2} homobilayer have reported an approximately tenfold increase in the nonlinearity of exciton dipolaritons compared to regular monolayer exciton-polaritons \cite{datta2022highly,louca}. Furthermore, moir\'e hybrid intra/interlayer exciton polaritons have also been reported in a twisted \ce{MoSe2}/\ce{WS2} heterobilayer \cite{zhang2021van}.

\begin{figure}[t]
    \centering
    \includegraphics[width=0.5\textwidth]{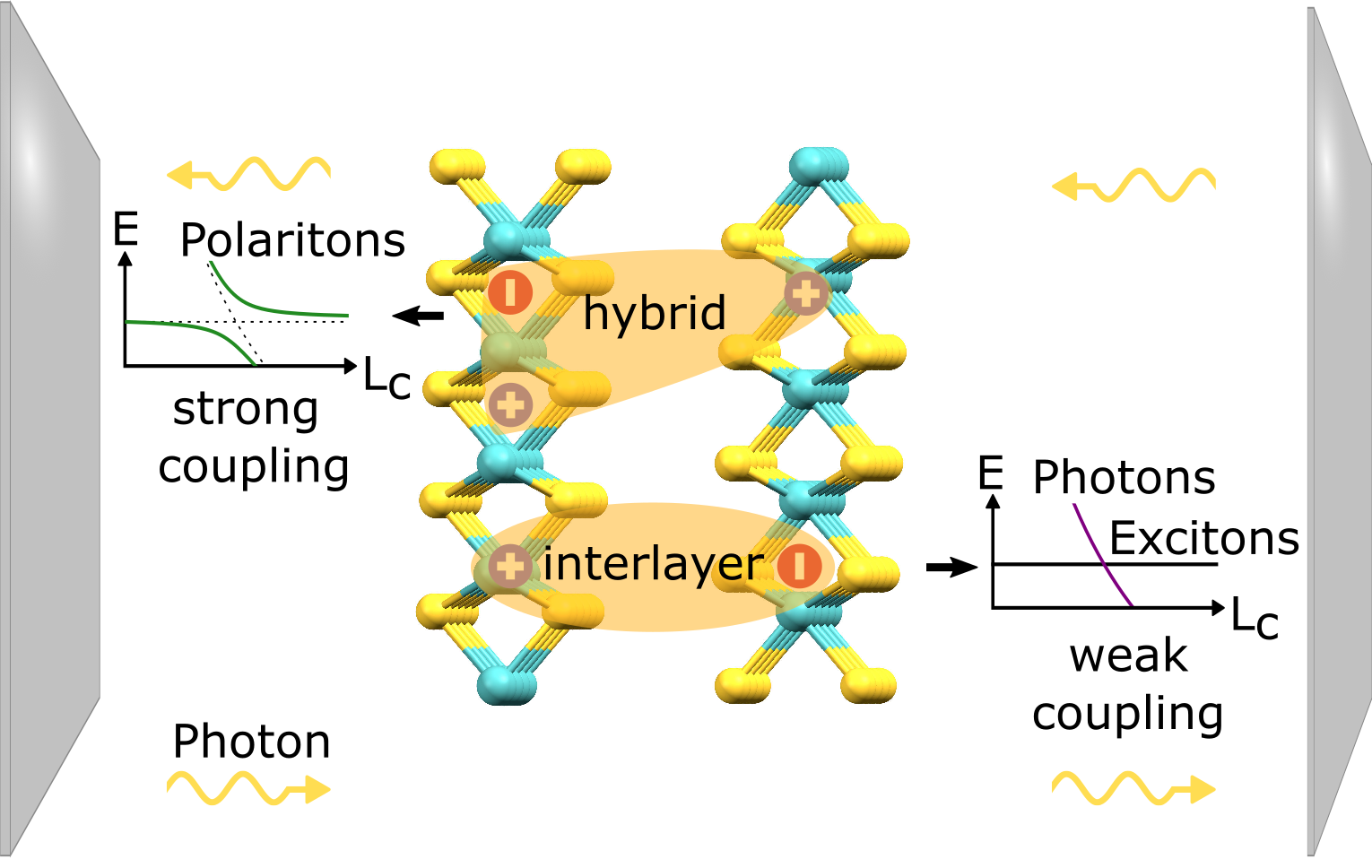}
    \caption{Schematic figure of $H_h^h$-stacked MoS$_2$ bilayer integrated within a Fabry-Perot cavity. While pure interlayer exciton show a very low oscillator strength and can only lead to a weak coupling regime with photons, hybridized excitons (that might consist to a large extent from interlayer excitons) can give rise to a strong coupling regime and formation of exciton polaritons. }
   \label{fig:schematic}
\end{figure}

Based on a microscopic and material-specific approach combining the density matrix formalism with the Hopfield theory, we demonstrate in this work that TMD homobilayers show clearly visible interlayer-like exciton polaritons in linear optical spectra. We focus our attention on the exemplary $H_h^h$-stacked \ce{MoS2} homobilayer integrated within a Fabry-Perot cavity (cf. Fig. \ref{fig:schematic}) due to its high degree of hybridization between layers. Remarkably, by carefully tuning the cavity length, it is possible to observe a middle polariton branch, where the excitonic contribution has an almost $100\%$ interlayer character, despite the extremely low oscillator strength of interlayer excitons. In a sense, the cavity photon can mediate an ``unmixing" of the constituent bare (i.e without hybridization) excitons. This means that despite the almost optically dark nature of the underlying interlayer component, the resulting polariton can still be readily observed in linear optical spectra. We explain this surprising finding in the context of the polariton absorption and critical coupling. In short, the intra/interlayer component of the polariton does not determine the optical properties, instead it is the photonic contribution and the relative weighting of radiative and exciton scattering decay rates \cite{Fitzgerald2022}. Our findings are of importance for the design of highly tunable and nonlinear exciton-based optoelectronic devices.

\section{Theoretical approach}
The energies and wavefunctions of the bare intralayer excitons (A, B) and interlayer excitons (IE$_1$, IE$_2$) are found by solving the bilayer Wannier equation \cite{ovesen2019interlayer}. Hereby, we focus on excitons with an optically allowed recombination and disregard spin- and momentum-dark states as they are not crucial in the strong coupling regime. In analogy to the A and B exciton, there are two energetically separated interlayer excitons. For H-type stacking, the energy difference is approximately given by the spin-orbit splitting of the valence and conduction band, cf. Figs. \ref{fig:hyb_exc}(a-b). The Hamiltonian, including the tunneling contribution, reads in the exciton basis \cite{hagel}
$\hat{H}_0=\sum_{L=1,\textbf{Q}}^4 E_{L,\textbf{Q}}^{(X)} \hat{X}_{L,\textbf{Q}}^\dagger \hat{X}_{L,\textbf{Q}}+\sum_{L\neq L',\bm{Q}}^4T_{LL'}\hat{X}_{L,\textbf{Q}}^\dagger \hat{X}_{L',\textbf{Q}},$
where $\hat{X}_{L,\textbf{Q}}^{(\dagger)}$ are the annihilation (creation) operators for the bare excitons with the center-of-mass momentum $\textbf{Q}$ and the layer configuration index $L$, which takes into account the four different bare excitons. Furthermore, $E_{L,\textbf{Q}}^{(X)}$ is the energy of the exciton and $T_{LL'}$ describes the tunneling matrix element. We consider only lattice-matched TMD bilayers without moiré effects. Furthermore, we focus on the lowest 1s energy exciton state of each layer configuration. The tunneling strength of the electron and hole can be obtained by DFT calculations and is taken from Ref. \citenum{hagel}. We consider hBN-encapsulated samples that are characterized by spectrally narrow exciton resonances. Thus, we assume a typical value of $1$ meV for the exciton scattering rate, which is appropriate for temperatures below $100$ K \cite{cadiz2017excitonic}. Although we predominately focus on the $H_h^h$-stacked \ce{MoS2} homobilayer in this work, as this is the material configuration with the highest degree of hybridization between the K points, we emphasise that our methods can be applied to other lattice-matched homo/heterobilayers. 

We diagonalize the Hamiltonian for each $\textbf{Q}$ in the hybrid-exciton basis with $\hat{Y}_{\eta,\bm{Q}} ^{(\dagger)}=\sum_{L=1}^4 \hat{X}_{L,\bm{Q}}^{(\dagger)} C_L^{\eta(*)}(\bm{Q})$ with $\hat{Y}_{\eta,\bm{Q}}^{(\dagger)}$ as annihilation (creation) operators for hybrid excitons. The new eigenstates are linear combinations of inter- and intralayer excitons determined by the tunneling between layers. The eigenvector $C_L^\eta$ quantifies the contributions of the $L$th bare exciton to the $\eta$th hybrid exciton (Fig. \ref{fig:hyb_exc}(a-b)). Depending on the degree of hybridization, the energy of excitons can be strongly shifted, as shown in  Fig. \ref{fig:hyb_exc}(c).

Next, we consider the coherent coupling between bilayer excitons and cavity photons to investigate whether the strong-coupling regime can be reached (Fig. \ref{fig:schematic}(a)). As we focus on normal incidence, we can drop the center-of-mass momentum dependence and the light-exciton Hamiltonian can be written in the hybrid basis within the rotating-wave approximation as 
\begin{align}
    \hat{H}&=\sum_{\eta=1}^4 E^{(Y)}_\eta \hat{Y}_\eta^\dagger \hat{Y}_\eta+ E^{(c)}\hat{c}^\dagger \hat{c}
    +\sum_{\eta=1}^4 g_\eta^{(Y)} \hat{Y}_\eta^\dagger \hat{c}+ \text{h.c.} \label{eq:Hamiltonian2}
\end{align}
with $\hat{c}^{(\dagger)}$ as the cavity photon field operators and $E^{(c)}$ as the corresponding photon energy. The transformed light-exciton coupling, $g_\eta^{(Y)}=\sum_{L=1}^4 C_L^\eta g_L^{(X)}$, describes how the oscillator strength of the bare excitons is shared out to the $\eta$th hybrid exciton. For the bare exciton-light coupling, we use a classically derived expression valid for the specific case of a thin excitonic material placed in the centre of a high-quality Fabry-Perot cavity close to resonance \cite{kavokin2003thin}, $g_\eta^{(X)}=\hbar\sqrt{\frac{\gamma_L}{\tau}\frac{1+|r_m|}{|r_m|}}$, where $r_m$ is the reflectivity of the mirrors (taken as $-0.99$ throughout the work), $\tau$ is the travel time of the photon inside the cavity, and $\hbar \gamma_L$ is the radiative coupling of the bare excitons \cite{kira2006many}.  

We use the Hopfield transformation to diagonalize Eq. \ref{eq:Hamiltonian2} and obtain  
\begin{align}
\hat{H}=\sum_{n=0}^4 E^{(P)}_n \hat{P}_n^\dagger \hat{P}_n,
\label{eq:Hamiltonian3}
\end{align}
where $\hat{P}_n^{(\dagger)}$ are the polariton field operators for the $n$th branch. This is achieved using the basis transformation $\hat{P}_n ^{(\dagger)}=\hat{c}^{(\dagger)}U_0^{n(*)}+\sum_{\eta=1}^4 \hat{Y}_\eta^{(\dagger)} U_\eta^{n(*)}$, where $U_\eta^{n(*)}$ are the Hopfield coefficients quantifying the contribution of the $\eta$th exciton/photon to the $n$th polariton. As there are two intralayer and two interlayer excitons (Fig. \ref{fig:hyb_exc}(a-b)), this leads to five polariton branches as we consider only the lowest-energy cavity mode (Fig. \ref{fig:length_sweep}). To provide insights into how the strong coupling regime modifies the mixing of intra- and interlayer excitons, it is instructive to transform the Hopfield coefficients back into the bare exciton basis. We name the resulting coefficients the \emph{generalized Hopfield coefficients}  
\begin{equation}
\tilde{U}_L^n=\sum_{\eta=1}^4 C_L^\eta U_\eta^n. \label{eq:generalised_Hopfield}    
\end{equation}
They quantify the contribution of the $L$th bare exciton to the $n$th polariton, i.e., they describe the intra/interlayer nature of an exciton dipolariton.

To make our predictions accessible in experiments, we link our microscopic model to a macroscopic observable and calculate the polariton absorption, which unambiguously demonstrates strong coupling physics via the Rabi splitting \cite{savona1995quantum}. To this end, we exploit an approximate input-output method \cite{collett1984squeezing} to calculate the polaritonic Elliot formula, which was recently introduced in Ref. \citenum{Fitzgerald2022}. This is valid for high-quality cavities and energetically well-spaced polaritons relative to their linewidth (which in practise will always be satisfied in the strong-coupling regime), and close to the cavity resonance. The polariton absorption reads for the $n$th branch
\begin{align}
\mathcal{A}_n(\omega)&=\frac{4\tilde{\gamma}_n\tilde{\Gamma}_n}{\left(\omega-E^{(P)}_n/\hbar\right)+(2\tilde{\gamma}_n + \tilde{\Gamma}_n)}\label{eq:polaritonic_elliot}
\end{align}
with the effective scattering rate $ \tilde{\Gamma}_n=(1-|U_{0}^n|^2)\Gamma$ and cavity leakage rate $
\tilde{\gamma}_n=\kappa|U_{0}^n|^2$ expressed in the polariton basis. 
Here, $U_{0}^n$ is the photonic Hopfield coefficient, $\kappa$ denotes the cavity leakage rate of a Fabry-Perot cavity, and $\Gamma$ is the bare exciton scattering rate. Crucially, this expression gives insight into how the excitonic and photonic nature of a polariton, as well as the balance between radiative and scattering decay channels, manifests in experimentally accessible absorption spectra \cite{Fitzgerald2022}.

\section{Results}
\subsection{Interlayer Exciton Polaritons}
\begin{figure}[t]
    \centering
    \includegraphics[width=0.5\textwidth]{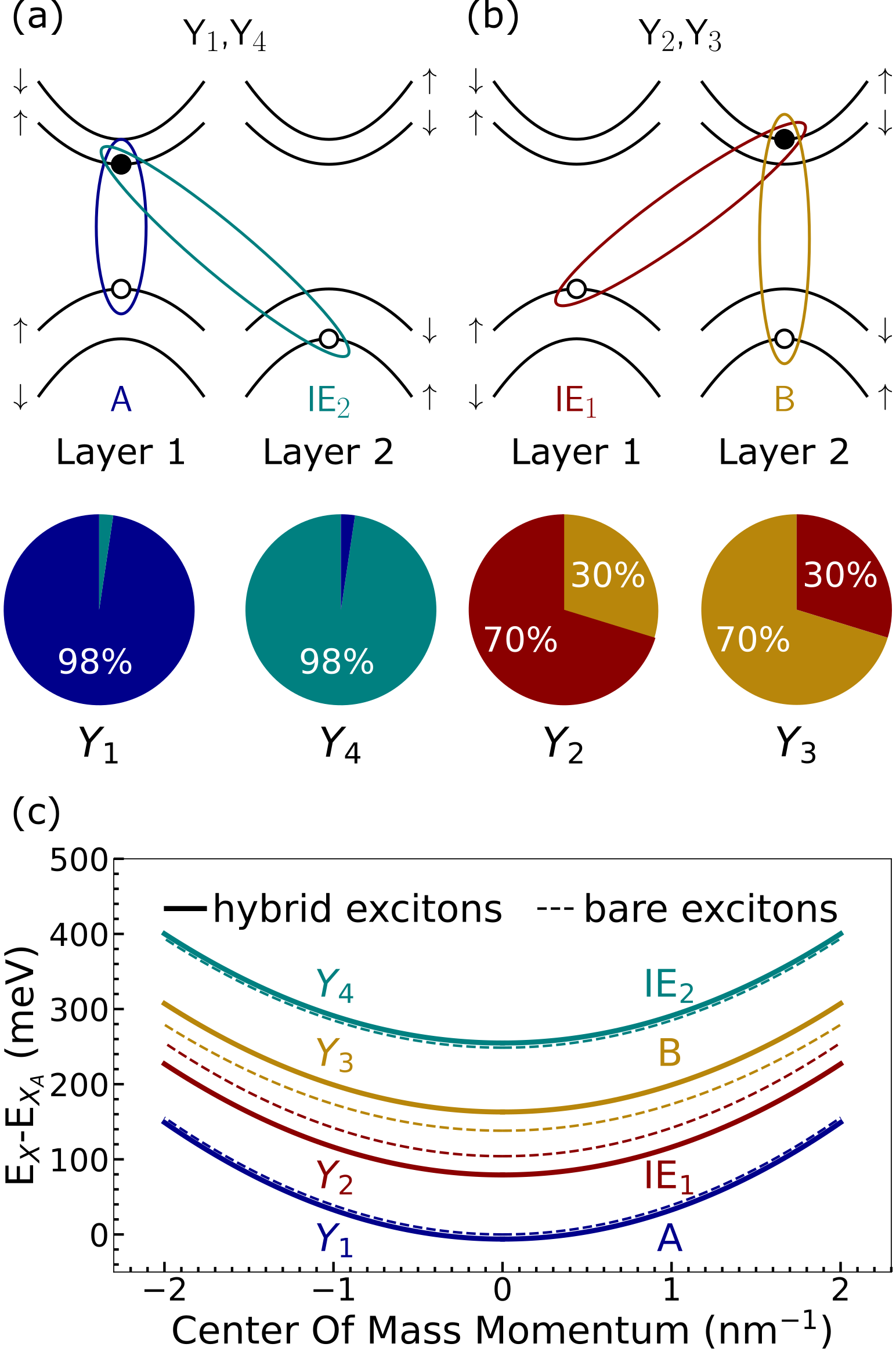}
    \caption{(a) The intralayer A and the interlayer IE2 exciton hybridize via hole tunneling to form the hybrid excitons Y$_1$ and Y$_4$. (b) The intralayer B and the interlayer IE1 exciton  hybridize via hole tunneling to form the hybrid excitons Y$_2$ and Y$_3$.
   (c) Energy of the bare (dashed lines) and hybridized (solid lines) excitons as a function of the center of mass momentum.}
   \label{fig:hyb_exc}
\end{figure}
We begin by briefly discussing the hybrid excitons of a bare TMD bilayer. The band configuration of the constituent bare excitons are shown in Figs. \ref{fig:hyb_exc}(a-b). We focus our study on an H$_h^h$-stacked homobilayer, where the transition-metal atoms of one layer are located above the chalcogenide atoms of the other, as this stacking tends to give the highest degree of hybridization \cite{hagel}. In this particular case, electron tunneling is forbidden due to the symmetry of the electronic wavefunctions, which are predominately composed of d-orbitals at the K-points of the Brillouin zone, so only hole tunneling is allowed \cite{hagel}. This decouples the hybridization into two parts (considering that only states with the same spin can hybridize): the A exciton couples only to the IE$_2$ interlayer exciton, and the B exciton only to IE$_1$. This can be also seen in the lower panel of Fig. \ref{fig:hyb_exc}(a-b): The hybrid excitons Y$_1$ and Y$_4$ consist of A and IE$_2$ only. Likewise, Y$_2$ and Y$_3$ are a hybridization of only the B and IE$_1$ excitons. This is further illustrated in an avoided crossing of the involved exciton states in Fig. \ref{fig:hyb_exc}(c). Crucially, despite the apparent equal energy separation between the A and IE$_2$ excitons, and B and IE$_1$ excitons expected from the electronic band picture presented in Figs. \ref{fig:hyb_exc} (a) and (b), the spacing, and consequently the hybridization, between these excitons is substantially different, cf. solid and dashed lines in  Fig. \ref{fig:hyb_exc} (c). This is due to the different binding energies of the exciton species and the flipped spin ordering of the layers in the H-type stacking. Further details are given in the SI. As a consequence, while Y$_1$ and Y$_4$ excitons are only weakly hybridized (they consist of either 98\% the A or the IE$_2$ exciton, cf. the lower panel in Fig. \ref{fig:hyb_exc}(a)),  Y$_2$ and Y$_3$ exhibit a strong hybridization. Given their large intralayer component ($30\%$ and $70\%$), Y$_2$ and Y$_3$ hybrid excitons are prime candidates for investigating the strong-coupling physics.

\begin{figure}[t]
    \centering
        \includegraphics[width=0.5\textwidth]{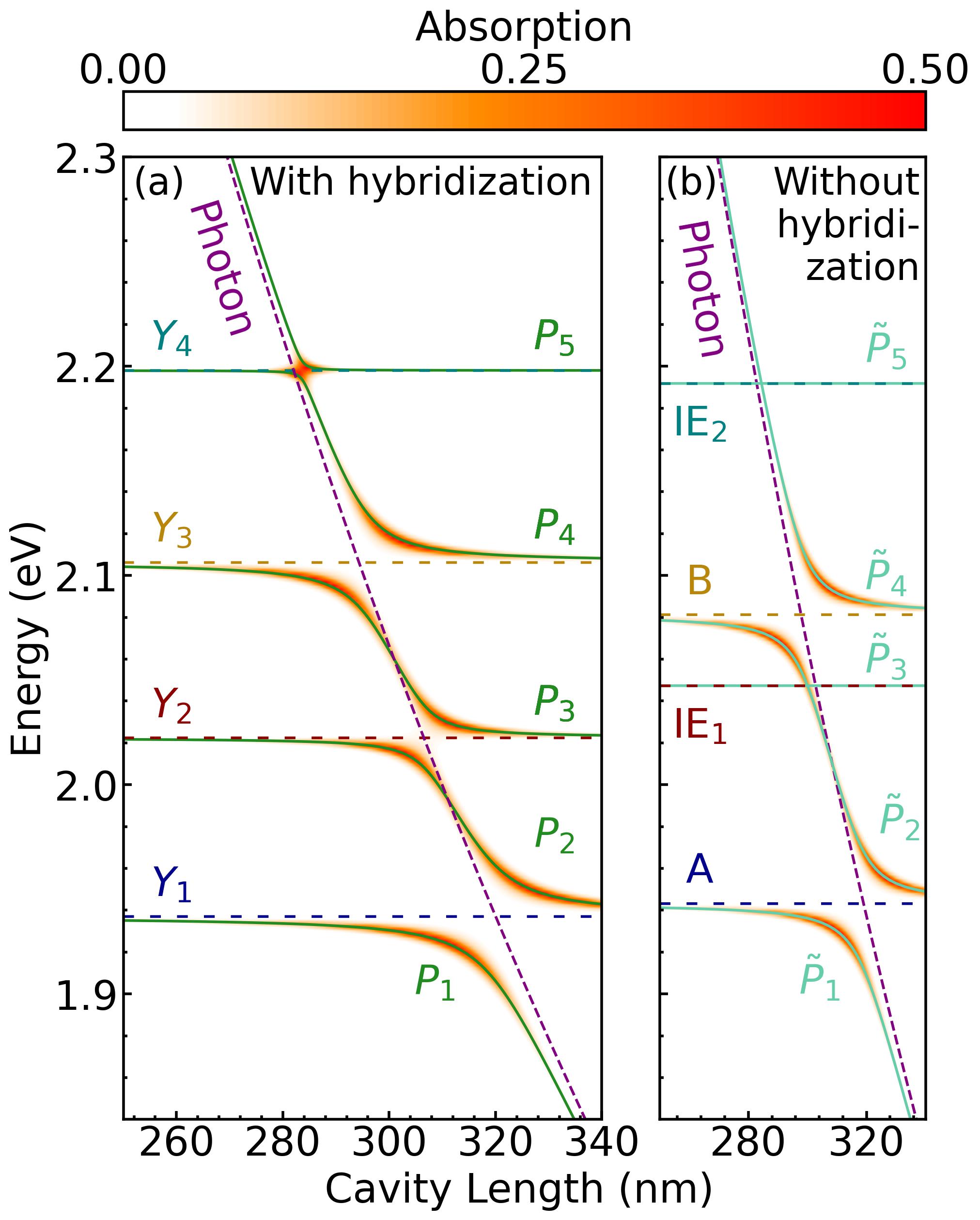}
    \caption{Length sweep of polariton energy (solid green lines) and absorption for a $H_h^h$-stacked MoS$_2$ bilayer integrated within the center of a Fabry-Perot microcavity with (a) and without (b) electronic hybridization. Cavity/exciton mode energy is given by the purple (blue) dashed line. Here, Y$_1$-Y$_4$ denote the energy of hybridized excitons, while P$_1$-P$_5$ and $\tilde{\mathrm{P}}_1$-$\tilde{\mathrm{P}}_5$ denote the different polariton branches with and without hybridization, respectively.}
    \label{fig:length_sweep}
\end{figure}

One means to explore the strong coupling regime is to detune exciton and photon energies by changing the cavity length \cite{Dufferwiel2015}. Figure \ref{fig:length_sweep} shows the polariton energy and absorption as a function of cavity length, with a vanishing in-plane photon momentum considered. To better understand the combined effects of electronic and photonic hybridization, we compare the polaritons of the full system (Fig. \ref{fig:length_sweep}(a)) with a fictitious system, where electronic tunneling is switched off (Fig. \ref{fig:length_sweep}(b)). In both cases, there are pronounced Rabi splittings of about $60$\,meV near to the A and B excitons energies corresponding to the formation of upper and lower polariton branches, but the hybrid exciton system shows additional two smaller splittings close to the bare interlayer exciton energies IE$_1$ and IE$_2$. This is a direct consequence of the exciton hybridization, with each polariton branch inheriting contributions from multiple hybrid excitons with the exact mixing depending on the cavity length. This in turn means that each polariton has a tunable mixed intra- and interlayer exciton nature, which can be understood via the light-exciton coupling $g^{(Y)}$. 
We compare the case of the hybridized exciton Y$_2$ and the pure interlayer exciton IE$_1$, which are close in energy. IE$_1$ has an oscillator strength approximately two orders of magnitude smaller than the intralayer excitons \cite{ovesen2019interlayer}, i.e. $g^{(X)}_2\approx 0$ and consequentially there is no Rabi splitting at this energy, cf. Fig. \ref{fig:length_sweep}(b). In contrast, Y$_2$ inherits a substantial intralayer component with $C_{B}^2=0.3$ and so  $g^{(Y)}_2\approx \sqrt{0.3}\, g^{(X)}_{B}\approx0.55\,g^{(X)}_{B}$, which is large enough to show a considerable Rabi splitting, cf. Fig. \ref{fig:length_sweep}(a). Note that exciton-light coupling varies with the cavity length, but this dependence is rather weak with about $4$\,meV change over the lengths considered in Fig. \ref{fig:length_sweep}.

To understand these results further, we employ the generalized Hopfield coefficients introduced in Eq. (\ref{eq:generalised_Hopfield}). We focus on the polaritons P$_1$ and P$_3$, which have a substantially different character, cf. Fig. \ref{fig:abs_ana} (a) and (d). A plot of the Hopfield coefficients (normal and generalized) for all branches can be found in the SI. The polariton P$_1$ behaves much like the lower polariton branch of the standard single-exciton single-photon model, i.e. as cavity length is increased, the polariton smoothly evolves from predominately A-exciton-like to photon-like (Fig. \ref{fig:abs_ana} (a)). The polariton  has only a very small interlayer component and this manifests as a slightly reduced and shifted Rabi splitting ($P$ vs $\tilde P$ in Figs. \ref{fig:length_sweep}(a) compared to (b)). In contrast, the polariton P$_3$ shows a much richer behaviour. While at small (large) cavity lengths, it limits to a fixed ratio of intralayer B and IE$_1$ excitons given by the composition of the hybrid exciton Y$_3$ (Y$_2$), at intermediate lengths it is a mix of intra-, interlayer excitons plus a photon component. This is an example of photon-induced hybridization of excitons \cite{Fitzgerald2022}, which occurs because the coupling strength $g^Y_\eta$ is comparable to the exciton energy separation. Interestingly, at cavity lengths near $300$ nm, the intralayer component is close to zero (Fig. \ref{fig:abs_ana} (d)), i.e. the hybridization with the photon 'unmixes' the initial hybridization between intra- and interlayer excitons as further discussed below. We denote this effect as \emph{photon-mediated dehybridization}.

Note that we obtain the same results when we first consider the hybridization between photons and bare excitons and then include the tunneling between different polaritons. In this picture, it is natural to think of a cavity-modified tunneling between polaritons \cite{cristofolini2012coupling}, i.e., a cavity-controllable charge transfer between TMD layers. In the SI, we show a cavity length study of this modified tunneling between polaritons.

\subsection{Critical Coupling for Polariton Absorption}

The results shown in Fig. \ref{fig:length_sweep} demonstrate that, despite the large interlayer component, the polariton P$_3$ exhibits a strong absorption, in fact, comparable to the predominately intralayer-like polariton P$_1$. Crucially, this indicates that polaritons with an interlayer character can be optically bright and hence directly visible in experiments. We explain this somewhat surprising result in the context of critical coupling and the polaritonic Elliot formula \cite{Fitzgerald2022}. In particular, Eq. (\ref{eq:polaritonic_elliot}) reveals that the maximum absorption is found when radiative and material-based losses are balanced, i.e., $2\tilde{\gamma}=\tilde{\Gamma}$. In the case of a mirror-symmetric two-port system that supports a single resonance, absorption is limited to a maximum of 50\% \cite{piper2014}. This polaritonic critical coupling condition depends purely on the photonic Hopfield coefficient $U_0^{n}$, and consequentially the intra- and interlayer percentage of the total \emph{excitonic} contribution does not directly impact the polariton absorption. Importantly, while a hybrid exciton needs a large intralayer component to enter the strong coupling regime in the first place, the resulting polariton need not have any intralayer contribution and, in principle, can possess a maximum possible absorption. 
\begin{figure}[t]
    \centering
    \includegraphics[width=0.5\textwidth]{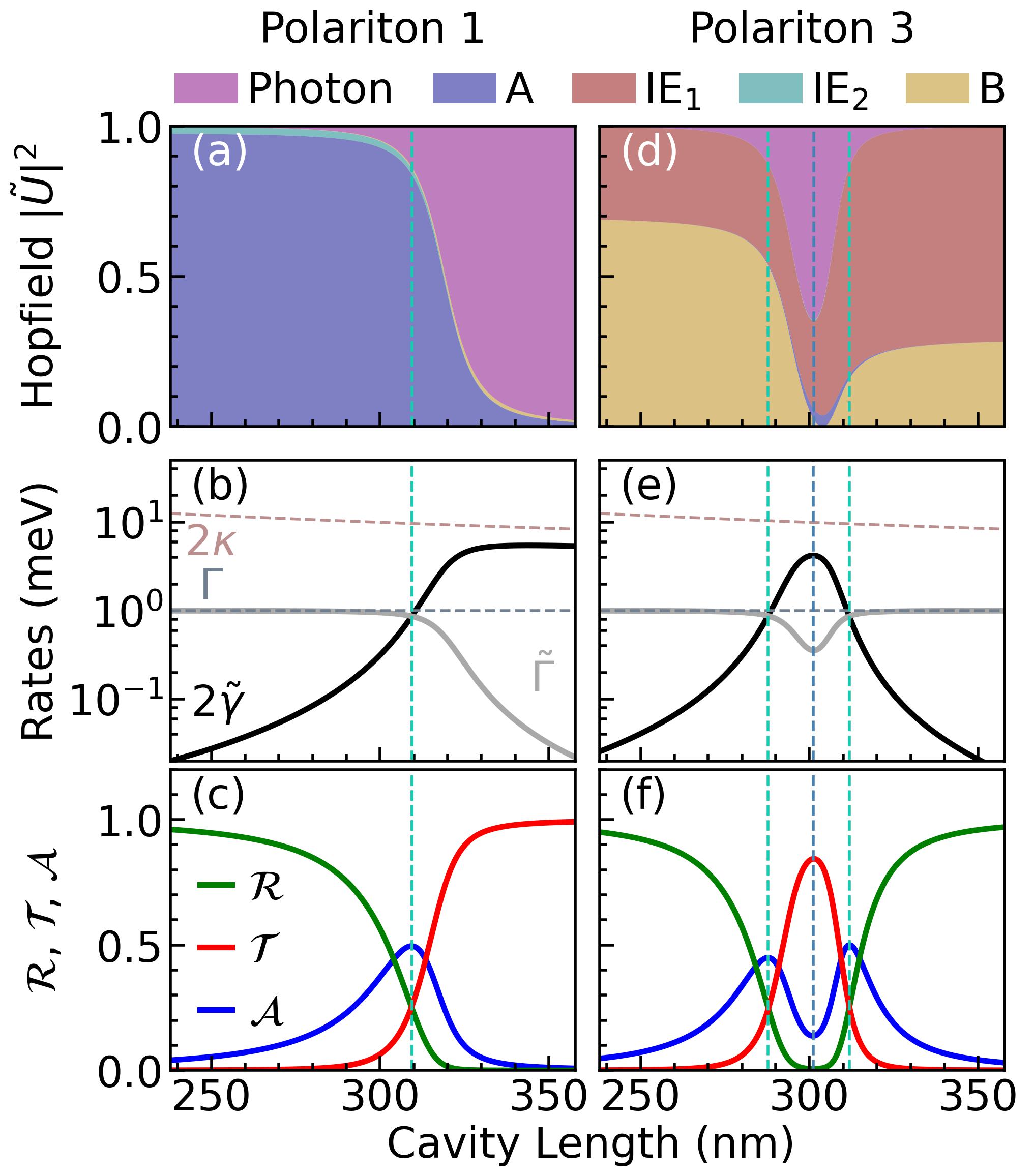}
    \caption{(a) The generalized Hopfield coefficients $\tilde{U}$ for the polariton $P_1$ as a function of the cavity length, where the turquoise vertical dashed line denotes the cavity length of maximum absorption. (b) Excitonic decay rate ($\tilde{\Gamma}$) and photonic decay rate ($2\tilde{\gamma}$)  as a function of cavity length for the polariton branch $P_1$. The crossing point of these lines denotes the critical coupling, where maximum absorption is reached. (c) Reflection, transmission and absorption as a function of cavity length for P$_1$.
    (d),(e),(f) Same as above but for the polariton branch P$_3$.} 
    \label{fig:abs_ana}
\end{figure}
This insight is confirmed in Fig. \ref{fig:abs_ana}, where we compare the linear optical spectra of the polariton P$_1$ and P$_3$ at resonance for different cavity lengths. For P$_1$,  we observe a typical behaviour for the reflection $\mathcal{R}$, transmission $\mathcal{T}$ and absorption $\mathcal{A}$ of the lower polariton branch. The absorption has a single peak corresponding to the critical coupling point, which is indicated in Fig. \ref{fig:abs_ana}(b), where the material loss (grey line, $\tilde{\Gamma}$) and the cavity loss (black line, $2\tilde{\gamma}$) lines cross. This is a consequence of the monotonic decrease in the photonic Hopfield coefficient for increasing length in Fig. \ref{fig:abs_ana}(a). In contrast, the polariton P$_3$ along with the other middle branches shows a double peak structure in absorption, cf.  Fig. \ref{fig:abs_ana}(f). This is a consequence of the critical coupling condition being met twice (Fig. \ref{fig:abs_ana}(e)) and can be traced back to the non-monotonic behaviour of the photonic Hopfield coefficient (Fig. \ref{fig:abs_ana}(d)). Interestingly, we also predict a local minimum in the absorption and a global maximum in the transmission, cf. the light blue vertical line. This line coincides with a local maximum in the difference between $\tilde{\Gamma}$ and $2\tilde{\gamma}$ (Fig. \ref{fig:abs_ana}(e)) and is a result of the maximum for the photonic Hopfield coefficient (Fig. \ref{fig:abs_ana}(d)). 

These results illustrate that the maximum in absorption is not dependent on the interlayer character of the polariton, but rather on the interplay between the photonic Hopfield coefficient and the relative magnitude of radiative and material-based loss (critical coupling). This means that polaritons with a very large interlayer exciton component can, in principle, possess a large signal in linear optical spectra.

\subsection{Photon-Mediated Dehybridization of Excitons}

It is insightful to introduce an interlayer contribution of the total exciton component for each polariton branch  $U_\mathrm{Inter}^n=(|U_{\mathrm{IE}_1}^n|^2+|U_{\mathrm{IE}_2}^n|^2)/(1-|U_{0}^n|^2)$ using the generalized Hopfield coefficients defined in Eq. \ref{eq:generalised_Hopfield}. In Fig. \ref{fig:inter_nonlinearity}(a), this quantity is shown for each polariton branch as a function of cavity length, with the color opacity proportional to the absorption at resonance. We find that only the polariton P$_3$ exhibits both an interlayer component maximum of $90\%$ and a significant absorption at the same cavity lengths. In contrast, the polaritons P$_4$ and $P_5$ become increasingly dark as they approach an almost 100$\%$ interlayer component at shorter and longer cavity lengths, respectively, i.e., the photonic Hopfield coefficient limits to zero. 

The polariton P$_3$ possesses this unique interlayer maximum due to a cancellation effect between the hybridized excitons Y$_2$ and Y$_3$. Both excitons have an inverted inter/intralayer contribution relative to one another (cf. Fig. \ref{fig:hyb_exc}(b)) and this leads to a photon-mediated cancellation of the intralayer component at a certain cavity length. It is actually the small contribution from the A exciton that prevents a total cancellation (cf. Fig. \ref{fig:abs_ana}(d)). To confirm this, we have introduced a gedanken experiment with a simplified two-exciton model where we consider only the two excitons IE$_1$ and B. In this case, the interlayer contribution of the middle polariton branch does indeed reach unity close to the interlayer maximum of P$_3$ (dotted line in Fig. \ref{fig:inter_nonlinearity}(a)). In the supplementary material, we demonstrate that for a suitable tuned cavity it is always possible to reach this interlayer maximum in the simplified two-exciton system. This rather non-intuitive behavior can be understood with the analogy of three linearly coupled harmonic oscillators, where the outer oscillators do not couple directly but only indirectly via the oscillator in between. This system, even in the case of asymmetric oscillators, can always support a normal mode where the middle oscillator is completely static for suitably tuned masses and spring constants. In our system the intralayer exciton plays the role of the middle oscillator and can mediate an interaction between the photon and interlayer exciton without absorbing any energy itself.

To further explore photon-mediated dehybridization in TMDs, we varied the tunneling strength of the hole in $H_h^h$-stacked \ce{MoS2} homobilayer. In particular, we are interested if it is possible to achieve a dipolariton with zero intralayer contribution, while remaining bright in absorption. While at first sights this seems a rather unrealistic study, it is possible to increase and reduce the interlayer coupling in van der Waals heterostructures using pressure and tensile strain, respectively \cite{zhu2022exchange}. Furthermore, the tunneling can be significantly reduced by adding hBN layers in between the TMD layers \cite{fang2014strong,calman2018indirect,sun2022excitonic}. We found that, while a higher tunneling strength increases the intralayer A exciton component of the middle polariton branch P$_3$, a lower tunneling strength can indeed give rise to a complete cancellation of the intralayer component (cf. the SI for more details). Crucially, this is disconnected from the absorption, which can be tuned to peak at any tunneling strength by changing the mirror reflectivity and exciton linewidths (critical coupling condition).

\subsection{Polariton-Polariton interaction}
\begin{figure}[t]
    \centering
    \includegraphics[width=0.5\textwidth]{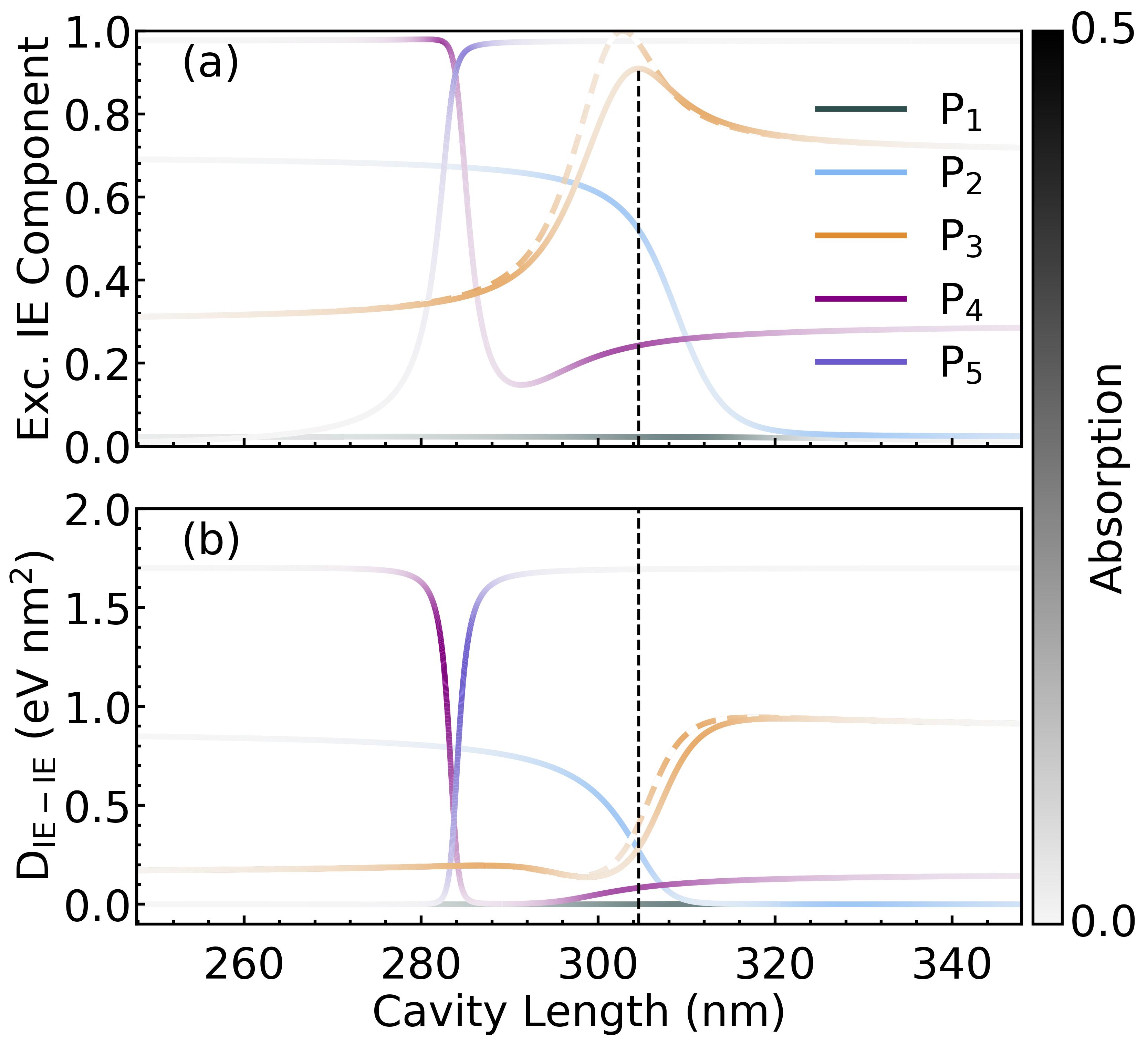}
    \caption{\textbf{(a)} Interlayer contribution of the total exciton component for the five polariton branches as a function of cavity length (solid lines). The dashed line indicates the corresponding result for P$_3$ within a simplified two-exciton model (as described in the text).
    The color gradient denotes the absorption intensity along the branches. \textbf{(b)} Dipolariton-dipolariton interaction for the different polariton branches as a function of cavity length.
    }
    \label{fig:inter_nonlinearity}
\end{figure}
It is interesting to explore the photon-induced dehybridization in the context of polariton-polariton interactions. Dipolaritons supported by TMD heterostructures offer promise for nonlinear optics in the low-photon density limit \cite{datta2022highly}, and our model can offer insight into their highly-tunable interactions. Polariton-polariton interactions involving intralayer exciton-polaritons have been extensively investigated on a microscopic footing \cite{PhysRevB.80.155306, PhysRevB.82.075301}. Here, we instead consider interactions between \emph{dipolaritons}, i.e., exciton-polaritons with a large interlayer exciton component. While intralayer excitons interact predominantly via exchange interactions \cite{ciuti1998role}, interlayer excitons couple via strong repulsive dipole-dipole interactions due to their permanent out-of-plane dipole moment \cite{erkensten2021exciton, li2020dipolar}. The dipole-dipole interaction of interlayer excitons constitutes the direct part of the full exciton-exciton coupling and can be derived from a bosonic exciton-exciton Hamiltonian \cite{erkensten2021exciton}. 

We consider the dipole-dipole interaction in the long wavelength limit given by $D^{\mathrm{IE}-\mathrm{IE}}\approx\frac{de^2}{\epsilon_0 \epsilon_{\perp}}$, corresponding to a contact potential in real space. Here, $d$ is the separation between the two TMD layers, approximately equal to the TMD layer thickness. Furthermore,  $\epsilon_\perp$  denotes the out-of-plane dielectric constant of the TMD layers. Dipolaritons are expected to partially inherit this enhanced scattering from the interlayer exciton \cite{datta2022highly, louca}. The repulsive dipole-dipole interaction leads to a blue-shifted emission spectrum for increasing polariton density $n_p$, where the blue-shift $\Delta E_n$ of the $n$th polariton branch can be quantified by the polariton-analogue of the widely used plate capacitor formula $\Delta E_n=D^n n_{p}$ \cite{schindler2008analysis} introducing the dipolariton-dipolariton interaction strength  
\begin{equation}
   D^n = D^{\mathrm{IE}-\mathrm{IE}}\left(\left|\tilde U_{\mathrm{IE}_1}^n\right|^4+\left|\tilde U_{\mathrm{IE}_2}^n\right|^4\right).\label{eq:nonlin} \ 
\end{equation}
The dipolariton-dipolariton interaction strength is obtained by applying the Hopfield transformation to the exciton-exciton interaction and is dictated by the interlayer \emph{generalized} Hopfield coefficients. Additional details on the derivation of the dipolariton-dipolariton interaction strength are found in the SI. 

Evaluating Eq. \ref{eq:nonlin}, we predict that the polaritons P$_4$ and P$_5$ show the largest nonlinearity in the limit of short or long cavities, respectively (cf. Fig. \ref{fig:inter_nonlinearity}(b)). However, these polaritons are almost optically dark at these cavity lengths (reflected by the vanishing opacity in Fig. \ref{fig:inter_nonlinearity}(b)) as they consist almost 100\% from the Y$_4$ exciton, which is very IE$_2$-like (cf. Figs. \ref{fig:length_sweep}(a) and \ref{fig:hyb_exc}(a)). Interestingly, the polaritons P$_2$ and P$_3$ are both bright and show a relatively high nonlinearity coefficient around the interlayer maximum of P$_3$ (black dashed line). Their photonic part is so large that the actual interlayer component is reasonably small (Fig. \ref{fig:abs_ana}(d)). In other words, while the interlayer component makes up a very large fraction of the total excitonic component of P$_3$, it is small in comparison to the photonic contribution. This leads to a substantially smaller dipole-dipole interaction compared to pure interlayer excitons. Nevertheless, both P$_2$ and P$_3$ still offer promise for nonlinear optical devices on the single-photon level due to their strong optical signatures and tunability. Note that we have here neglected the saturation nonlinearity, i.e., the density dependent Rabi splitting, which has been recently suggested to be comparable in magnitude to dipole mediated nonlinearity at low polariton densities \cite{datta2022highly}, as this is beyond the scope of this work.

One means to boost the nonlinearity of the middle polariton branches is to enhance the exciton-light interaction. Somewhat counter-intuitively, excitons that are close in energy, relative to the coupling strength, can be strongly mixed by the cavity photon without the resulting middle polariton branches possessing a large photonic component \cite{Fitzgerald2022}. This means that a larger coupling leads to flatter polariton bands with a lower photonic contribution, which in turn increases the dipolariton-dipolariton interaction strength. For a given exciton scattering rate, the critical coupling condition dictates that polaritons with a small photonic Hopfield coefficient can still be visible in linear optical spectra for a suitable increase in the cavity decay rate. This suggests that plasmonic cavities, which have demonstrated Rabi splittings in the hundreds of meV \cite{wang2016coherent}, could offer advantages for nonlinear polaritonics.

\section{Conclusions}
Based on a microscopic and material-realistic model, we predict clear optical signatures of interlayer exciton polaritons in $H_h^h$-stacked \ce{MoS2} homobilayers integrated within a Fabry-Perot cavity. This occurs despite the very low oscillator strength of interlayer excitons and has its microscopic origin in the strong hybridization with intralayer excitons. Furthermore, we have shown that photon-mediated hybridization can significantly unmix the initial hybridization between intra- and interlayer excitons, as the exciton-photon coupling strength is comparable to the exciton energy separation. This results in the formation of bright polaritons with near 100\% interlayer excitonic character. 
While we have focused on cavity length as a dial to control interlayer exciton polaritons, further experimentally accessible knobs for tuning are provided by electric field \cite{peimyoo2021electrical}, temperature \cite{epstein2020near}, strain \cite{niehues2019interlayer} and twist angle \cite{zhang2021van,Fitzgerald2022}. Overall, combining electronic and photonic hybridization offers a new strategy to efficiently couple optical energy into dark material-based excitations, which could be of importance in exciton-based optoelectronics devices.

Supplementary information accompanies the manuscript on the Light:
Science \& Applications website (http://www.nature.com/lsa)


\begin{acknowledgements}
We acknowledge funding from the European Union’s Horizon 2020 research and innovation program under Grant Agreement No. 881603 (Graphene Flagship), the DFG via SFB 1083 (Project B9), and the Knut and Alice Wallenberg Foundation (2014.0226).
\end{acknowledgements}

\bibliography{ref}

\begin{thebibliography}{10}
\expandafter\ifx\csname url\endcsname\relax
  \def\url#1{\texttt{#1}}\fi
\expandafter\ifx\csname urlprefix\endcsname\relax\def\urlprefix{URL }\fi
\providecommand{\bibinfo}[2]{#2}
\providecommand{\eprint}[2][]{\url{#2}}

\bibitem{chernikov2014exciton}
\bibinfo{author}{Chernikov, A.} \emph{et~al.}
\newblock \bibinfo{title}{Exciton binding energy and nonhydrogenic rydberg
  series in monolayer ws 2}.
\newblock \emph{\bibinfo{journal}{Physical review letters}}
  \textbf{\bibinfo{volume}{113}}, \bibinfo{pages}{076802}
  (\bibinfo{year}{2014}).

\bibitem{perea2022exciton}
\bibinfo{author}{Perea-Causin, R.} \emph{et~al.}
\newblock \bibinfo{title}{Exciton optics, dynamics, and transport in atomically
  thin semiconductors}.
\newblock \emph{\bibinfo{journal}{APL Materials}}
  \textbf{\bibinfo{volume}{10}}, \bibinfo{pages}{100701}
  (\bibinfo{year}{2022}).

\bibitem{rivera2015observation}
\bibinfo{author}{Rivera, P.} \emph{et~al.}
\newblock \bibinfo{title}{Observation of long-lived interlayer excitons in
  monolayer mose2--wse2 heterostructures}.
\newblock \emph{\bibinfo{journal}{Nature communications}}
  \textbf{\bibinfo{volume}{6}}, \bibinfo{pages}{1--6} (\bibinfo{year}{2015}).

\bibitem{calman2018indirect}
\bibinfo{author}{Calman, E.} \emph{et~al.}
\newblock \bibinfo{title}{Indirect excitons in van der waals heterostructures
  at room temperature}.
\newblock \emph{\bibinfo{journal}{Nature communications}}
  \textbf{\bibinfo{volume}{9}}, \bibinfo{pages}{1--5} (\bibinfo{year}{2018}).

\bibitem{jiang2021interlayer}
\bibinfo{author}{Jiang, Y.}, \bibinfo{author}{Chen, S.},
  \bibinfo{author}{Zheng, W.}, \bibinfo{author}{Zheng, B.} \&
  \bibinfo{author}{Pan, A.}
\newblock \bibinfo{title}{Interlayer exciton formation, relaxation, and
  transport in tmd van der waals heterostructures}.
\newblock \emph{\bibinfo{journal}{Light: Science \& Applications}}
  \textbf{\bibinfo{volume}{10}}, \bibinfo{pages}{1--29} (\bibinfo{year}{2021}).

\bibitem{ovesen2019interlayer}
\bibinfo{author}{Ovesen, S.} \emph{et~al.}
\newblock \bibinfo{title}{Interlayer exciton dynamics in van der waals
  heterostructures}.
\newblock \emph{\bibinfo{journal}{Communications Physics}}
  \textbf{\bibinfo{volume}{2}}, \bibinfo{pages}{1--8} (\bibinfo{year}{2019}).

\bibitem{merkl2019ultrafast}
\bibinfo{author}{Merkl, P.} \emph{et~al.}
\newblock \bibinfo{title}{Ultrafast transition between exciton phases in van
  der waals heterostructures}.
\newblock \emph{\bibinfo{journal}{Nature materials}}
  \textbf{\bibinfo{volume}{18}}, \bibinfo{pages}{691--696}
  (\bibinfo{year}{2019}).

\bibitem{lorchat2021excitons}
\bibinfo{author}{Lorchat, E.} \emph{et~al.}
\newblock \bibinfo{title}{Excitons in bilayer mos 2 displaying a colossal
  electric field splitting and tunable magnetic response}.
\newblock \emph{\bibinfo{journal}{Physical Review Letters}}
  \textbf{\bibinfo{volume}{126}}, \bibinfo{pages}{037401}
  (\bibinfo{year}{2021}).

\bibitem{krasnok2018nanophotonics}
\bibinfo{author}{Krasnok, A.}, \bibinfo{author}{Lepeshov, S.} \&
  \bibinfo{author}{Al\'{u}, A.}
\newblock \bibinfo{title}{Nanophotonics with 2d transition metal
  dichalcogenides}.
\newblock \emph{\bibinfo{journal}{Optics express}}
  \textbf{\bibinfo{volume}{26}}, \bibinfo{pages}{15972--15994}
  (\bibinfo{year}{2018}).

\bibitem{schneider2018two}
\bibinfo{author}{Schneider, C.}, \bibinfo{author}{Glazov, M.~M.},
  \bibinfo{author}{Korn, T.}, \bibinfo{author}{H{\"o}fling, S.} \&
  \bibinfo{author}{Urbaszek, B.}
\newblock \bibinfo{title}{Two-dimensional semiconductors in the regime of
  strong light-matter coupling}.
\newblock \emph{\bibinfo{journal}{Nature communications}}
  \textbf{\bibinfo{volume}{9}}, \bibinfo{pages}{1--9} (\bibinfo{year}{2018}).

\bibitem{kavokin2003thin}
\bibinfo{author}{Kavokin, A.} \& \bibinfo{author}{Malpuech, G.}
\newblock In \emph{\bibinfo{booktitle}{Cavity Polaritons}}, Thin films and
  Nanostructures, \bibinfo{pages}{29--45} (\bibinfo{publisher}{Elsevier},
  \bibinfo{year}{2003}).

\bibitem{Dufferwiel2015}
\bibinfo{author}{Dufferwiel, S.} \emph{et~al.}
\newblock \bibinfo{title}{Exciton–polaritons in van der waals
  heterostructures embedded in tunable microcavities}.
\newblock \emph{\bibinfo{journal}{Nature Communications}}
  \textbf{\bibinfo{volume}{6}}, \bibinfo{pages}{8579} (\bibinfo{year}{2015}).
\newblock \urlprefix\url{https://doi.org/10.1038/ncomms9579}.

\bibitem{liu2016strong}
\bibinfo{author}{Liu, W.} \emph{et~al.}
\newblock \bibinfo{title}{Strong exciton--plasmon coupling in mos2 coupled with
  plasmonic lattice}.
\newblock \emph{\bibinfo{journal}{Nano letters}} \textbf{\bibinfo{volume}{16}},
  \bibinfo{pages}{1262--1269} (\bibinfo{year}{2016}).

\bibitem{latini2019cavity}
\bibinfo{author}{Latini, S.}, \bibinfo{author}{Ronca, E.},
  \bibinfo{author}{De~Giovannini, U.}, \bibinfo{author}{H{\"u}bener, H.} \&
  \bibinfo{author}{Rubio, A.}
\newblock \bibinfo{title}{Cavity control of excitons in two-dimensional
  materials}.
\newblock \emph{\bibinfo{journal}{Nano letters}} \textbf{\bibinfo{volume}{19}},
  \bibinfo{pages}{3473--3479} (\bibinfo{year}{2019}).

\bibitem{Fitzgerald2022}
\bibinfo{author}{Fitzgerald, J.~M.}, \bibinfo{author}{Thompson, J. J.~P.} \&
  \bibinfo{author}{Malic, E.}
\newblock \bibinfo{title}{Twist angle tuning of moiré exciton polaritons in
  van der waals heterostructures}.
\newblock \emph{\bibinfo{journal}{Nano Letters}} \textbf{\bibinfo{volume}{22}},
  \bibinfo{pages}{4468--4474} (\bibinfo{year}{2022}).
\newblock \urlprefix\url{https://doi.org/10.1021/acs.nanolett.2c01175}.
\newblock \bibinfo{note}{PMID: 35594200},
  \eprint{https://doi.org/10.1021/acs.nanolett.2c01175}.

\bibitem{ross2017interlayer}
\bibinfo{author}{Ross, J.~S.} \emph{et~al.}
\newblock \bibinfo{title}{Interlayer exciton optoelectronics in a 2d
  heterostructure p--n junction}.
\newblock \emph{\bibinfo{journal}{Nano letters}} \textbf{\bibinfo{volume}{17}},
  \bibinfo{pages}{638--643} (\bibinfo{year}{2017}).

\bibitem{forg2019cavity}
\bibinfo{author}{F{\"o}rg, M.} \emph{et~al.}
\newblock \bibinfo{title}{Cavity-control of interlayer excitons in van der
  waals heterostructures}.
\newblock \emph{\bibinfo{journal}{Nature communications}}
  \textbf{\bibinfo{volume}{10}}, \bibinfo{pages}{1--6} (\bibinfo{year}{2019}).

\bibitem{miller2017long}
\bibinfo{author}{Miller, B.} \emph{et~al.}
\newblock \bibinfo{title}{Long-lived direct and indirect interlayer excitons in
  van der waals heterostructures}.
\newblock \emph{\bibinfo{journal}{Nano letters}} \textbf{\bibinfo{volume}{17}},
  \bibinfo{pages}{5229--5237} (\bibinfo{year}{2017}).

\bibitem{nagler2017interlayer}
\bibinfo{author}{Nagler, P.} \emph{et~al.}
\newblock \bibinfo{title}{Interlayer exciton dynamics in a dichalcogenide
  monolayer heterostructure}.
\newblock \emph{\bibinfo{journal}{2D Materials}} \textbf{\bibinfo{volume}{4}},
  \bibinfo{pages}{025112} (\bibinfo{year}{2017}).

\bibitem{erkensten2021exciton}
\bibinfo{author}{Erkensten, D.}, \bibinfo{author}{Brem, S.} \&
  \bibinfo{author}{Malic, E.}
\newblock \bibinfo{title}{Exciton-exciton interaction in transition metal
  dichalcogenide monolayers and van der waals heterostructures}.
\newblock \emph{\bibinfo{journal}{Physical Review B}}
  \textbf{\bibinfo{volume}{103}}, \bibinfo{pages}{045426}
  (\bibinfo{year}{2021}).

\bibitem{unuchek2018room}
\bibinfo{author}{Unuchek, D.} \emph{et~al.}
\newblock \bibinfo{title}{Room-temperature electrical control of exciton flux
  in a van der waals heterostructure}.
\newblock \emph{\bibinfo{journal}{Nature}} \textbf{\bibinfo{volume}{560}},
  \bibinfo{pages}{340--344} (\bibinfo{year}{2018}).

\bibitem{jauregui2019electrical}
\bibinfo{author}{Jauregui, L.~A.} \emph{et~al.}
\newblock \bibinfo{title}{Electrical control of interlayer exciton dynamics in
  atomically thin heterostructures}.
\newblock \emph{\bibinfo{journal}{Science}} \textbf{\bibinfo{volume}{366}},
  \bibinfo{pages}{870--875} (\bibinfo{year}{2019}).

\bibitem{leisgang2020giant}
\bibinfo{author}{Leisgang, N.} \emph{et~al.}
\newblock \bibinfo{title}{Giant stark splitting of an exciton in bilayer mos2}.
\newblock \emph{\bibinfo{journal}{Nature Nanotechnology}}
  \textbf{\bibinfo{volume}{15}}, \bibinfo{pages}{901--907}
  (\bibinfo{year}{2020}).

\bibitem{peimyoo2021electrical}
\bibinfo{author}{Peimyoo, N.} \emph{et~al.}
\newblock \bibinfo{title}{Electrical tuning of optically active interlayer
  excitons in bilayer mos2}.
\newblock \emph{\bibinfo{journal}{Nature Nanotechnology}}
  \textbf{\bibinfo{volume}{16}}, \bibinfo{pages}{888--893}
  (\bibinfo{year}{2021}).

\bibitem{hagel2022electrical}
\bibinfo{author}{Hagel, J.}, \bibinfo{author}{Brem, S.} \&
  \bibinfo{author}{Malic, E.}
\newblock \bibinfo{title}{Electrical tuning of moir$\backslash$'e excitons in
  mose $ \_2 $ bilayers}.
\newblock \emph{\bibinfo{journal}{arXiv preprint arXiv:2207.01890}}
  (\bibinfo{year}{2022}).

\bibitem{fang2014strong}
\bibinfo{author}{Fang, H.} \emph{et~al.}
\newblock \bibinfo{title}{Strong interlayer coupling in van der waals
  heterostructures built from single-layer chalcogenides}.
\newblock \emph{\bibinfo{journal}{Proceedings of the National Academy of
  Sciences}} \textbf{\bibinfo{volume}{111}}, \bibinfo{pages}{6198--6202}
  (\bibinfo{year}{2014}).

\bibitem{alexeev2019resonantly}
\bibinfo{author}{Alexeev, E.~M.} \emph{et~al.}
\newblock \bibinfo{title}{Resonantly hybridized excitons in moir{\'e}
  superlattices in van der waals heterostructures}.
\newblock \emph{\bibinfo{journal}{Nature}} \textbf{\bibinfo{volume}{567}},
  \bibinfo{pages}{81--86} (\bibinfo{year}{2019}).

\bibitem{brem2020hybridized}
\bibinfo{author}{Brem, S.} \emph{et~al.}
\newblock \bibinfo{title}{Hybridized intervalley moir{\'e} excitons and flat
  bands in twisted wse 2 bilayers}.
\newblock \emph{\bibinfo{journal}{Nanoscale}} \textbf{\bibinfo{volume}{12}},
  \bibinfo{pages}{11088--11094} (\bibinfo{year}{2020}).

\bibitem{gerber2019interlayer}
\bibinfo{author}{Gerber, I.~C.} \emph{et~al.}
\newblock \bibinfo{title}{Interlayer excitons in bilayer mos 2 with strong
  oscillator strength up to room temperature}.
\newblock \emph{\bibinfo{journal}{Physical Review B}}
  \textbf{\bibinfo{volume}{99}}, \bibinfo{pages}{035443}
  (\bibinfo{year}{2019}).

\bibitem{cristofolini2012coupling}
\bibinfo{author}{Cristofolini, P.} \emph{et~al.}
\newblock \bibinfo{title}{Coupling quantum tunneling with cavity photons}.
\newblock \emph{\bibinfo{journal}{Science}} \textbf{\bibinfo{volume}{336}},
  \bibinfo{pages}{704--707} (\bibinfo{year}{2012}).

\bibitem{rosenberg2018strongly}
\bibinfo{author}{Rosenberg, I.} \emph{et~al.}
\newblock \bibinfo{title}{Strongly interacting dipolar-polaritons}.
\newblock \emph{\bibinfo{journal}{Science Advances}}
  \textbf{\bibinfo{volume}{4}}, \bibinfo{pages}{eaat8880}
  (\bibinfo{year}{2018}).

\bibitem{datta2022highly}
\bibinfo{author}{Datta, B.} \emph{et~al.}
\newblock \bibinfo{title}{Highly nonlinear dipolar exciton-polaritons in
  bilayer mos2}.
\newblock \emph{\bibinfo{journal}{Nature Communications}}
  \textbf{\bibinfo{volume}{13}}, \bibinfo{pages}{6341} (\bibinfo{year}{2022}).
\newblock \urlprefix\url{https://doi.org/10.1038/s41467-022-33940-3}.

\bibitem{louca}
\bibinfo{author}{Louca, C.} \emph{et~al.}
\newblock \bibinfo{title}{Nonlinear interactions of dipolar excitons and
  polaritons in mos2 bilayers}.
\newblock \emph{\bibinfo{journal}{arXiv preprint arXiv:2204.00485}}
  (\bibinfo{year}{2022}).

\bibitem{zhang2021van}
\bibinfo{author}{Zhang, L.} \emph{et~al.}
\newblock \bibinfo{title}{Van der waals heterostructure polaritons with
  moir{\'e}-induced nonlinearity}.
\newblock \emph{\bibinfo{journal}{Nature}} \textbf{\bibinfo{volume}{591}},
  \bibinfo{pages}{61--65} (\bibinfo{year}{2021}).

\bibitem{hagel}
\bibinfo{author}{Hagel, J.}, \bibinfo{author}{Brem, S.},
  \bibinfo{author}{Linder{\"a}lv, C.}, \bibinfo{author}{Erhart, P.} \&
  \bibinfo{author}{Malic, E.}
\newblock \bibinfo{title}{Exciton landscape in van der waals heterostructures}.
\newblock \emph{\bibinfo{journal}{Physical Review Research}}
  \textbf{\bibinfo{volume}{3}}, \bibinfo{pages}{043217} (\bibinfo{year}{2021}).

\bibitem{cadiz2017excitonic}
\bibinfo{author}{Cadiz, F.} \emph{et~al.}
\newblock \bibinfo{title}{Excitonic linewidth approaching the homogeneous limit
  in mos 2-based van der waals heterostructures}.
\newblock \emph{\bibinfo{journal}{Physical Review X}}
  \textbf{\bibinfo{volume}{7}}, \bibinfo{pages}{021026} (\bibinfo{year}{2017}).

\bibitem{kira2006many}
\bibinfo{author}{Kira, M.} \& \bibinfo{author}{Koch, S.~W.}
\newblock \bibinfo{title}{Many-body correlations and excitonic effects in
  semiconductor spectroscopy}.
\newblock \emph{\bibinfo{journal}{Progress in quantum electronics}}
  \textbf{\bibinfo{volume}{30}}, \bibinfo{pages}{155--296}
  (\bibinfo{year}{2006}).

\bibitem{savona1995quantum}
\bibinfo{author}{Savona, V.}, \bibinfo{author}{Andreani, L.},
  \bibinfo{author}{Schwendimann, P.} \& \bibinfo{author}{Quattropani, A.}
\newblock \bibinfo{title}{Quantum well excitons in semiconductor microcavities:
  Unified treatment of weak and strong coupling regimes}.
\newblock \emph{\bibinfo{journal}{Solid State Communications}}
  \textbf{\bibinfo{volume}{93}}, \bibinfo{pages}{733--739}
  (\bibinfo{year}{1995}).

\bibitem{collett1984squeezing}
\bibinfo{author}{Collett, M.} \& \bibinfo{author}{Gardiner, C.}
\newblock \bibinfo{title}{Squeezing of intracavity and traveling-wave light
  fields produced in parametric amplification}.
\newblock \emph{\bibinfo{journal}{Physical Review A}}
  \textbf{\bibinfo{volume}{30}}, \bibinfo{pages}{1386} (\bibinfo{year}{1984}).

\bibitem{piper2014}
\bibinfo{author}{Piper, J.~R.}, \bibinfo{author}{Liu, V.} \&
  \bibinfo{author}{Fan, S.}
\newblock \bibinfo{title}{Total absorption by degenerate critical coupling}.
\newblock \emph{\bibinfo{journal}{Appl. Phys. Lett.}}
  \textbf{\bibinfo{volume}{104}}, \bibinfo{pages}{251110}
  (\bibinfo{year}{2014}).

\bibitem{zhu2022exchange}
\bibinfo{author}{Zhu, M.} \emph{et~al.}
\newblock \bibinfo{title}{Exchange between interlayer and intralayer exciton in
  wse2/ws2 heterostructure by interlayer coupling engineering}.
\newblock \emph{\bibinfo{journal}{Nano Letters}}  (\bibinfo{year}{2022}).

\bibitem{sun2022excitonic}
\bibinfo{author}{Sun, Z.} \emph{et~al.}
\newblock \bibinfo{title}{Excitonic transport driven by repulsive dipolar
  interaction in a van der waals heterostructure}.
\newblock \emph{\bibinfo{journal}{Nature Photonics}}
  \textbf{\bibinfo{volume}{16}}, \bibinfo{pages}{79--85}
  (\bibinfo{year}{2022}).

\bibitem{PhysRevB.80.155306}
\bibinfo{author}{Glazov, M.~M.} \emph{et~al.}
\newblock \bibinfo{title}{Polariton-polariton scattering in microcavities: A
  microscopic theory}.
\newblock \emph{\bibinfo{journal}{Phys. Rev. B}} \textbf{\bibinfo{volume}{80}},
  \bibinfo{pages}{155306} (\bibinfo{year}{2009}).

\bibitem{PhysRevB.82.075301}
\bibinfo{author}{Vladimirova, M.} \emph{et~al.}
\newblock \bibinfo{title}{Polariton-polariton interaction constants in
  microcavities}.
\newblock \emph{\bibinfo{journal}{Phys. Rev. B}} \textbf{\bibinfo{volume}{82}},
  \bibinfo{pages}{075301} (\bibinfo{year}{2010}).

\bibitem{ciuti1998role}
\bibinfo{author}{Ciuti, C.}, \bibinfo{author}{Savona, V.},
  \bibinfo{author}{Piermarocchi, C.}, \bibinfo{author}{Quattropani, A.} \&
  \bibinfo{author}{Schwendimann, P.}
\newblock \bibinfo{title}{Role of the exchange of carriers in elastic
  exciton-exciton scattering in quantum wells}.
\newblock \emph{\bibinfo{journal}{Physical Review B}}
  \textbf{\bibinfo{volume}{58}}, \bibinfo{pages}{7926} (\bibinfo{year}{1998}).

\bibitem{li2020dipolar}
\bibinfo{author}{Li, W.}, \bibinfo{author}{Lu, X.}, \bibinfo{author}{Dubey,
  S.}, \bibinfo{author}{Devenica, L.} \& \bibinfo{author}{Srivastava, A.}
\newblock \bibinfo{title}{Dipolar interactions between localized interlayer
  excitons in van der waals heterostructures}.
\newblock \emph{\bibinfo{journal}{Nature Materials}}
  \textbf{\bibinfo{volume}{19}}, \bibinfo{pages}{624--629}
  (\bibinfo{year}{2020}).

\bibitem{schindler2008analysis}
\bibinfo{author}{Schindler, C.} \& \bibinfo{author}{Zimmermann, R.}
\newblock \bibinfo{title}{Analysis of the exciton-exciton interaction in
  semiconductor quantum wells}.
\newblock \emph{\bibinfo{journal}{Physical Review B}}
  \textbf{\bibinfo{volume}{78}}, \bibinfo{pages}{045313}
  (\bibinfo{year}{2008}).

\bibitem{wang2016coherent}
\bibinfo{author}{Wang, S.} \emph{et~al.}
\newblock \bibinfo{title}{Coherent coupling of ws2 monolayers with metallic
  photonic nanostructures at room temperature}.
\newblock \emph{\bibinfo{journal}{Nano letters}} \textbf{\bibinfo{volume}{16}},
  \bibinfo{pages}{4368--4374} (\bibinfo{year}{2016}).

\bibitem{epstein2020near}
\bibinfo{author}{Epstein, I.} \emph{et~al.}
\newblock \bibinfo{title}{Near-unity light absorption in a monolayer ws2 van
  der waals heterostructure cavity}.
\newblock \emph{\bibinfo{journal}{Nano Letters}} \textbf{\bibinfo{volume}{20}},
  \bibinfo{pages}{3545--3552} (\bibinfo{year}{2020}).

\bibitem{niehues2019interlayer}
\bibinfo{author}{Niehues, I.}, \bibinfo{author}{Blob, A.},
  \bibinfo{author}{Stiehm, T.}, \bibinfo{author}{de~Vasconcellos, S.~M.} \&
  \bibinfo{author}{Bratschitsch, R.}
\newblock \bibinfo{title}{Interlayer excitons in bilayer mos 2 under uniaxial
  tensile strain}.
\newblock \emph{\bibinfo{journal}{Nanoscale}} \textbf{\bibinfo{volume}{11}},
  \bibinfo{pages}{12788--12792} (\bibinfo{year}{2019}).

\end{thebibliography}


\begin{thebibliography}{10}
\expandafter\ifx\csname url\endcsname\relax
  \def\url#1{\texttt{#1}}\fi
\expandafter\ifx\csname urlprefix\endcsname\relax\def\urlprefix{URL }\fi
\providecommand{\bibinfo}[2]{#2}
\providecommand{\eprint}[2][]{\url{#2}}

\bibitem{Kormanyos}
\bibinfo{author}{Korm{\'{a}}nyos, A.} \emph{et~al.}
\newblock \bibinfo{title}{Corrigendum: k.p theory for two-dimensional
  transition metal dichalcogenide semiconductors (2015 2d mater. 2 022001)}.
\newblock \emph{\bibinfo{journal}{2D Materials}} \textbf{\bibinfo{volume}{2}},
  \bibinfo{pages}{049501} (\bibinfo{year}{2015}).
\newblock \urlprefix\url{https://doi.org/10.1088/2053-1583/2/4/049501}.

\bibitem{laturia2018dielectric}
\bibinfo{author}{Laturia, A.}, \bibinfo{author}{Van~de Put, M.~L.} \&
  \bibinfo{author}{Vandenberghe, W.~G.}
\newblock \bibinfo{title}{Dielectric properties of hexagonal boron nitride and
  transition metal dichalcogenides: from monolayer to bulk}.
\newblock \emph{\bibinfo{journal}{npj 2D Materials and Applications}}
  \textbf{\bibinfo{volume}{2}}, \bibinfo{pages}{1--7} (\bibinfo{year}{2018}).

\bibitem{hagel}
\bibinfo{author}{Hagel, J.}, \bibinfo{author}{Brem, S.},
  \bibinfo{author}{Linder{\"a}lv, C.}, \bibinfo{author}{Erhart, P.} \&
  \bibinfo{author}{Malic, E.}
\newblock \bibinfo{title}{Exciton landscape in van der waals heterostructures}.
\newblock \emph{\bibinfo{journal}{Physical Review Research}}
  \textbf{\bibinfo{volume}{3}}, \bibinfo{pages}{043217} (\bibinfo{year}{2021}).

\bibitem{li2014measurement}
\bibinfo{author}{Li, Y.} \emph{et~al.}
\newblock \bibinfo{title}{Measurement of the optical dielectric function of
  monolayer transition-metal dichalcogenides: Mos 2, mo s e 2, ws 2, and ws e
  2}.
\newblock \emph{\bibinfo{journal}{Physical Review B}}
  \textbf{\bibinfo{volume}{90}}, \bibinfo{pages}{205422}
  (\bibinfo{year}{2014}).

\bibitem{louca}
\bibinfo{author}{Louca, C.} \emph{et~al.}
\newblock \bibinfo{title}{Nonlinear interactions of dipolar excitons and
  polaritons in mos2 bilayers}.
\newblock \emph{\bibinfo{journal}{arXiv preprint arXiv:2204.00485}}
  (\bibinfo{year}{2022}).

\bibitem{ovesen2019interlayer}
\bibinfo{author}{Ovesen, S.} \emph{et~al.}
\newblock \bibinfo{title}{Interlayer exciton dynamics in van der waals
  heterostructures}.
\newblock \emph{\bibinfo{journal}{Communications Physics}}
  \textbf{\bibinfo{volume}{2}}, \bibinfo{pages}{1--8} (\bibinfo{year}{2019}).

\bibitem{Fitzgerald2022}
\bibinfo{author}{Fitzgerald, J.~M.}, \bibinfo{author}{Thompson, J. J.~P.} \&
  \bibinfo{author}{Malic, E.}
\newblock \bibinfo{title}{Twist angle tuning of moiré exciton polaritons in
  van der waals heterostructures}.
\newblock \emph{\bibinfo{journal}{Nano Letters}} \textbf{\bibinfo{volume}{22}},
  \bibinfo{pages}{4468--4474} (\bibinfo{year}{2022}).
\newblock \urlprefix\url{https://doi.org/10.1021/acs.nanolett.2c01175}.
\newblock \bibinfo{note}{PMID: 35594200},
  \eprint{https://doi.org/10.1021/acs.nanolett.2c01175}.

\bibitem{vasilevskiy2015exciton}
\bibinfo{author}{Vasilevskiy, M.~I.}, \bibinfo{author}{Santiago-Perez, D.~G.},
  \bibinfo{author}{Trallero-Giner, C.}, \bibinfo{author}{Peres, N.~M.} \&
  \bibinfo{author}{Kavokin, A.}
\newblock \bibinfo{title}{Exciton polaritons in two-dimensional dichalcogenide
  layers placed in a planar microcavity: Tunable interaction between two
  bose-einstein condensates}.
\newblock \emph{\bibinfo{journal}{Physical Review B}}
  \textbf{\bibinfo{volume}{92}}, \bibinfo{pages}{245435}
  (\bibinfo{year}{2015}).

\bibitem{zhu2022exchange}
\bibinfo{author}{Zhu, M.} \emph{et~al.}
\newblock \bibinfo{title}{Exchange between interlayer and intralayer exciton in
  wse2/ws2 heterostructure by interlayer coupling engineering}.
\newblock \emph{\bibinfo{journal}{Nano Letters}}  (\bibinfo{year}{2022}).

\bibitem{erkensten2021exciton}
\bibinfo{author}{Erkensten, D.}, \bibinfo{author}{Brem, S.} \&
  \bibinfo{author}{Malic, E.}
\newblock \bibinfo{title}{Exciton-exciton interaction in transition metal
  dichalcogenide monolayers and van der waals heterostructures}.
\newblock \emph{\bibinfo{journal}{Physical Review B}}
  \textbf{\bibinfo{volume}{103}}, \bibinfo{pages}{045426}
  (\bibinfo{year}{2021}).

\bibitem{kyriienko2012spin}
\bibinfo{author}{Kyriienko, O.}, \bibinfo{author}{Magnusson, E.} \&
  \bibinfo{author}{Shelykh, I.~A.}
\newblock \bibinfo{title}{Spin dynamics of cold exciton condensates}.
\newblock \emph{\bibinfo{journal}{Physical Review B}}
  \textbf{\bibinfo{volume}{86}}, \bibinfo{pages}{115324}
  (\bibinfo{year}{2012}).

\bibitem{erkensten2022microscopic}
\bibinfo{author}{Erkensten, D.}, \bibinfo{author}{Brem, S.},
  \bibinfo{author}{Perea-Caus{\'\i}n, R.} \& \bibinfo{author}{Malic, E.}
\newblock \bibinfo{title}{Microscopic origin of anomalous interlayer exciton
  transport in van der waals heterostructures}.
\newblock \emph{\bibinfo{journal}{Physical Review Materials}}
  \textbf{\bibinfo{volume}{6}}, \bibinfo{pages}{094006} (\bibinfo{year}{2022}).

\bibitem{brem2020hybridized}
\bibinfo{author}{Brem, S.} \emph{et~al.}
\newblock \bibinfo{title}{Hybridized intervalley moir{\'e} excitons and flat
  bands in twisted wse 2 bilayers}.
\newblock \emph{\bibinfo{journal}{Nanoscale}} \textbf{\bibinfo{volume}{12}},
  \bibinfo{pages}{11088--11094} (\bibinfo{year}{2020}).

\end{thebibliography}

\end{document}


\title{Supplementary Information\\Interlayer exciton polaritons in homobilayers of transition metal dichalcogenides}
\author{Jonas K. König}
\email{jonas.koenig@physik.uni-marburg.de}
\affiliation{Fachbereich Physik, Philipps-Universit\"at, Marburg, 35032, Germany}
\author{Jamie M. Fitzgerald}
\affiliation{Fachbereich Physik, Philipps-Universit\"at, Marburg, 35032, Germany}
\affiliation{Department of Physics, Chalmers University of Technology, SE-412 96 Gothenburg, Sweden}
\author{Joakim Hagel}
\affiliation{Department of Physics, Chalmers University of Technology, SE-412 96 Gothenburg, Sweden}
\author{Daniel Erkensten}
\affiliation{Department of Physics, Chalmers University of Technology, SE-412 96 Gothenburg, Sweden}
\author{Ermin Malic}
\affiliation{Fachbereich Physik, Philipps-Universit\"at, Marburg, 35032, Germany}
\affiliation{Department of Physics, Chalmers University of Technology, SE-412 96 Gothenburg, Sweden}
\maketitle
\section{Methods and input parameters}

Linear optical spectra have been calculated within the input-output formalism resulting in  the polaritonic Elliot formula presented in equation (4) of the main text. All results have been benchmarked against the exact transfer-matrix method and there is excellent agreement close to the cavity resonance.

The DFT parameters used throughout this work are taken from Refs. \citenum{Kormanyos} and \citenum{laturia2018dielectric}, while the tunneling strengths for the electronic hybridization are obtained from Ref. \citenum{hagel}. The optical matrix elements of the intralayer excitons are set to their monolayer value from a fit to experimental data \cite{li2014measurement}, and for the interlayer excitons set to zero. The band gap was set such that the energy of the hybrid exciton Y$_1$ matches the experimental value given in Ref. \citenum{louca}. Then, Y$_2$ and Y$_3$ align within 10 and 20 meV to their respective experimental values in the aforementioned reference.

\section{Exciton energy landscape}
Our analysis of exciton energy landscape is based on solving the Wannier equation, where we take into account the screening due to the dielectric environment \cite{ovesen2019interlayer}. Here, we focus on excitons consisting of carriers with the same spin and momentum. In a homobilayer, while only considering s-type excitons at the $K$ point, there are a total of $2^4=16$ different combinations of electrons and holes due to the spin splitting of each valence (VB) and conduction band (CB). These can be split into eight degenerate pairs with opposite spin configuration. Of these remaining eight excitons, four are spin-dark, and so we only explicitly consider four excitons, cf. Figs. 2(a)-(b) in the main text. Two of them are interlayer (IE$_1$ and IE$_2$) and two intralayer (A and B). The dispersion of these four excitons can be seen in Fig. 2(c) in the main text.

\begin{figure}[t]
    \centering
    \includegraphics[width=0.85\textwidth]{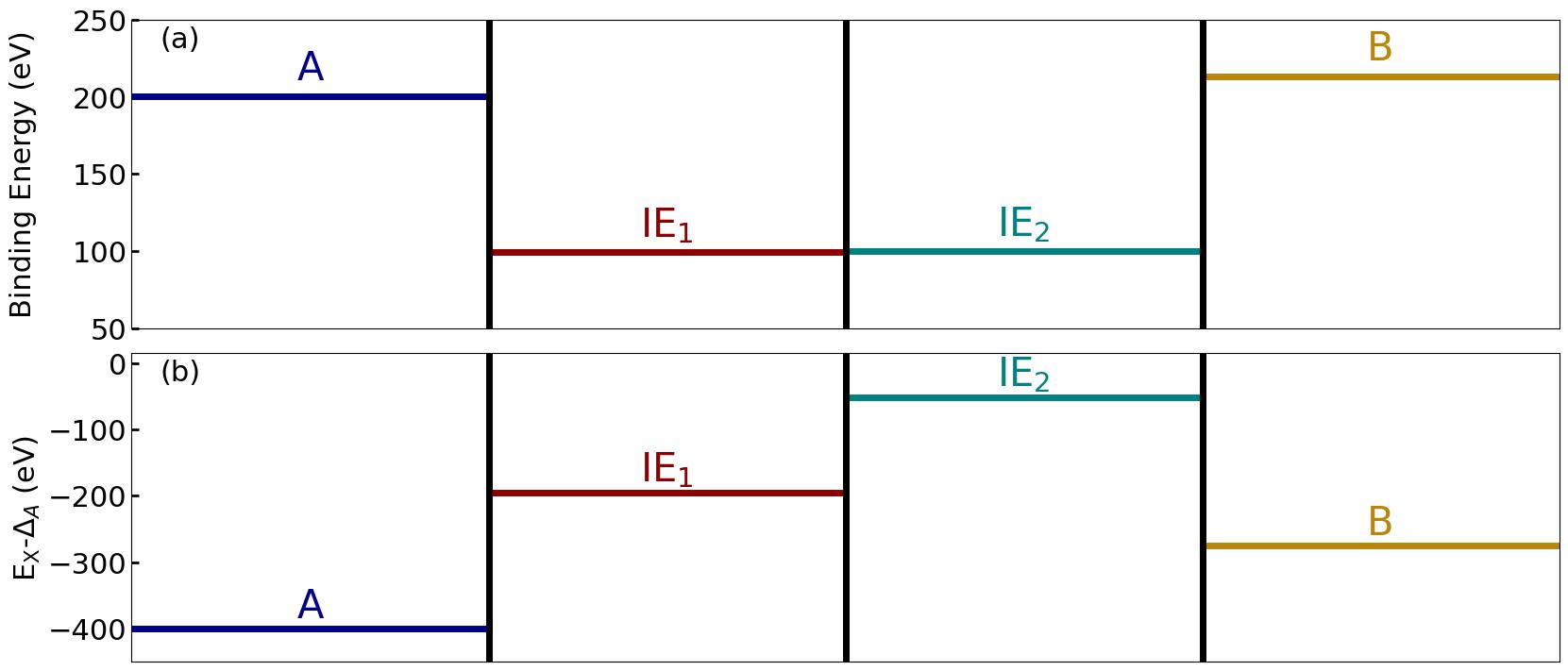}
    \caption{(a) Binding energies for the four considered bare excitons. (b) Spectral position of the above bare excitons relative to the band gap of the A exciton.}
    \label{fig:landscape}
\end{figure}

Here, we explain why the degree of hybridization as well as the energy shift relative to the bare excitons is much higher for Y$_2$ and Y$_3$ compared to Y$_1$ and Y$_4$. This is despite the splitting in the electronic band picture shown in Fig. 2(a) of the main text appearing to be equivalent in both cases. To this end, we have to explain why the B and IE$_1$ excitons are energetically closer compared to the A and IE$_2$ excitons. First we look at the different band gaps. If we relate every band gap to the band gap of the A exciton we obtain
$
    \Delta_{\mathrm{IE}_1}=\Delta_\mathrm{A}+\delta_C\;\text{,}\;
    \Delta_{\mathrm{IE}_2}=\Delta_\mathrm{A}+\delta_V\;\text{,}\;
    \Delta_{\mathrm{B}}=\Delta_\mathrm{A}+\delta_V+\delta_C\;\text{,}$ where $\delta_V$ and $\delta_C$ denote the spin-orbit splitting of the VB and CB, respectively (cf. Figs. 2(a)-(b) in the main text). In addition, the total energy of an exciton depends on its binding energy. If we assume the binding energies of both of the intralayer excitons to be approximately the same, and similarly for the interlayer excitons, we can relate all coupling exciton energies together as 
\begin{align*}
    |E_{\mathrm{IE}_2}-E_\mathrm{A}|&=|\delta_V+E^b_\mathrm{Intra}-E^b_\mathrm{Inter}|\;\text{,}\\
    |E_{\mathrm{IE}_1}-E_\mathrm{B}|&=|\delta_V-(E^b_\mathrm{Intra}-E^b_\mathrm{Inter})|\;\text{,}
\end{align*} 
where $E^b_\mathrm{Intra}$ and $E^b_\mathrm{Inter}$ denote the binding energies of intra- and interlayer excitons, respectively. The difference in the energy separation is thus given by two times the difference between inter- and intralayer binding energies. In the case of an MoS$_2$ homobilayer, the CB splitting is practically negligible compared to the VB splitting. Furthermore, the VB splitting is in the same order of magnitude as the difference in the binding energies between intra- and interlayer excitons. This then means that we can approximately write $|E_{\mathrm{IE}_2}-E_\mathrm{A}|\approx2\delta_V$ and $|E_{\mathrm{IE}_1}-E_\mathrm{B}|\approx0$. In other words, the B exciton is energetically much closer to the IE$_1$ than the A exciton is to the IE$_2$. The binding energies as well as the the spectral position of the excitons relative to the band gap of the A exciton are shown in Fig. \ref{fig:landscape}.

\section{Combining photonic and electronic hybridization}
\begin{figure}[t]
    \centering
    \includegraphics[width=1\textwidth]{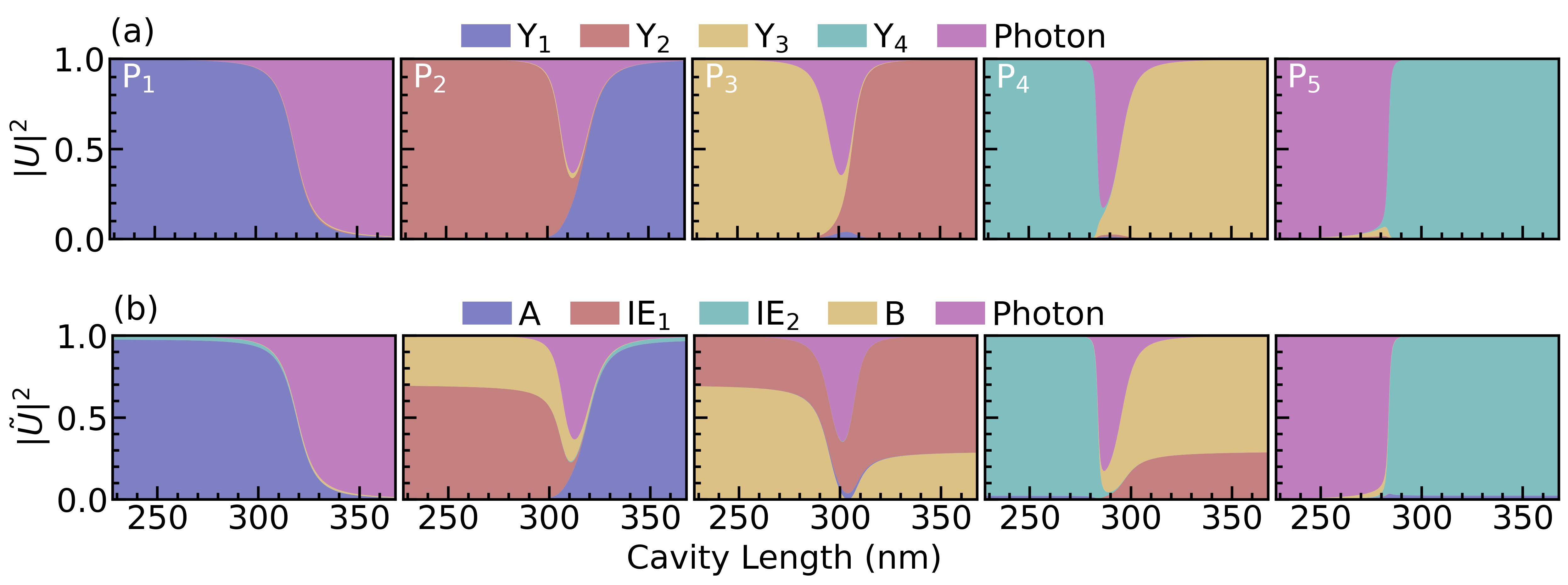}
    \caption{(a) Normal and (b) generalized Hopfield coefficients for all polariton branches as a function of cavity length.}
    \label{fig:All_hop}
\end{figure}

The employed Hopfield method has already been thoroughly described in the supplementary information of Ref. \citenum{Fitzgerald2022}. Here,  we discuss the extension to include electronic hybridisation. The complete Hamiltonian in the exciton basis is given by
\begin{equation}
    \hat{H}=\sum_{L=1}^4 E_L^{(X)} \hat{X}_L^\dagger \hat{X}_L+\sum_{L\neq L'}^4T_{LL'}\hat{X}_L^\dagger \hat{X}_{L'}+E^{(c)}\hat{c}^\dagger \hat{c}+\sum_L^4 g_L^{(X)} \hat{X}^\dagger \hat{c}+ \text{h.c.}\label{eq:full_hamiltonian}
\end{equation}
with $\hat{c}^{(\dagger)}$ denoting the annihilation (creation) operator and $E^{(c)}$  the energy of the cavity photon, $X^{(\dagger)}$ denoting the annihilation (creation) operator and $E_L^{(X)}$ the energy of the excitons, while $g_L^{(X)}$ describes the coupling between the cavity photon and the bare exciton with the layer index $L$. There are two ways to diagonalize this Hamiltonian: (1) the approach described in the main text where we transform first into the hybrid exciton basis and then diagonalize it using the Hopfield method, or (2) we transform it first into the exciton polariton basis and then take into account electronic hybridisation. Both methods, of course, have to give the same results, but since tunneling can not be switched off in reality, it is more intuitive to take approach (1). All standard and generalized Hopfield coefficients obtained with method (1) are shown in Fig. \ref{fig:All_hop}.

\section{Tunneling polaritons}
Here, we discuss method (2) from above. First, we transform Eq. (\ref{eq:full_hamiltonian}) into the polariton basis. In particular, the interlayer electronic tunneling term transforms as
\begin{subequations}
\begin{align}
    \hat{H}&=\sum_{nn'}\left(\tilde T_{nn'}+\delta_{nn'}\right) \hat{\tilde P}_n^\dagger \hat{\tilde P}_{n'}\;\text{,}\label{eq:hophyb}\\
    \tilde T_{nn'}&=\sum_{LL'}U_{L'}^{n'*}T_{LL'}U_{L}^{n}\;\text{.}\label{eq:new_tunnel}
\end{align}
\end{subequations}
where $\tilde P^{(\dagger)}$ denotes the non-hybridized exciton polariton annihilation (creation) operator. $\tilde T_{nn'}$ describes the tunneling between interlayer and intralayer exciton polaritons. The diagonal elements of this new matrix renormalize the energy of the different non-hybridized exciton polaritons. Here, these are always zero, as there is only tunneling between interlayer and intralayer excitons, but not directly between two interlayer or two intralayer excitons, i.e. $\tilde T_{nn'}$ is only non-zero if the $n$th polariton branch has an interlayer component, while the $n'$th has a component of the corresponding coupling intralayer exciton, and vice versa. As we set the oscillator strength of the interlayer excitons to zero, there are only two non-hybridized polariton branches with an interlayer component ($\tilde P_3$ and $\tilde P_5$ in Fig. (2) of the main text). This means that from the total 25 matrix elements (stemming from five different polaritons) only twelve are nonzero, which are six complex conjugate pairs as $\tilde T_{nn'}^\dagger=\tilde T_{nn'}$.

\begin{figure}[t]
    \centering
    \includegraphics[width=1\textwidth]{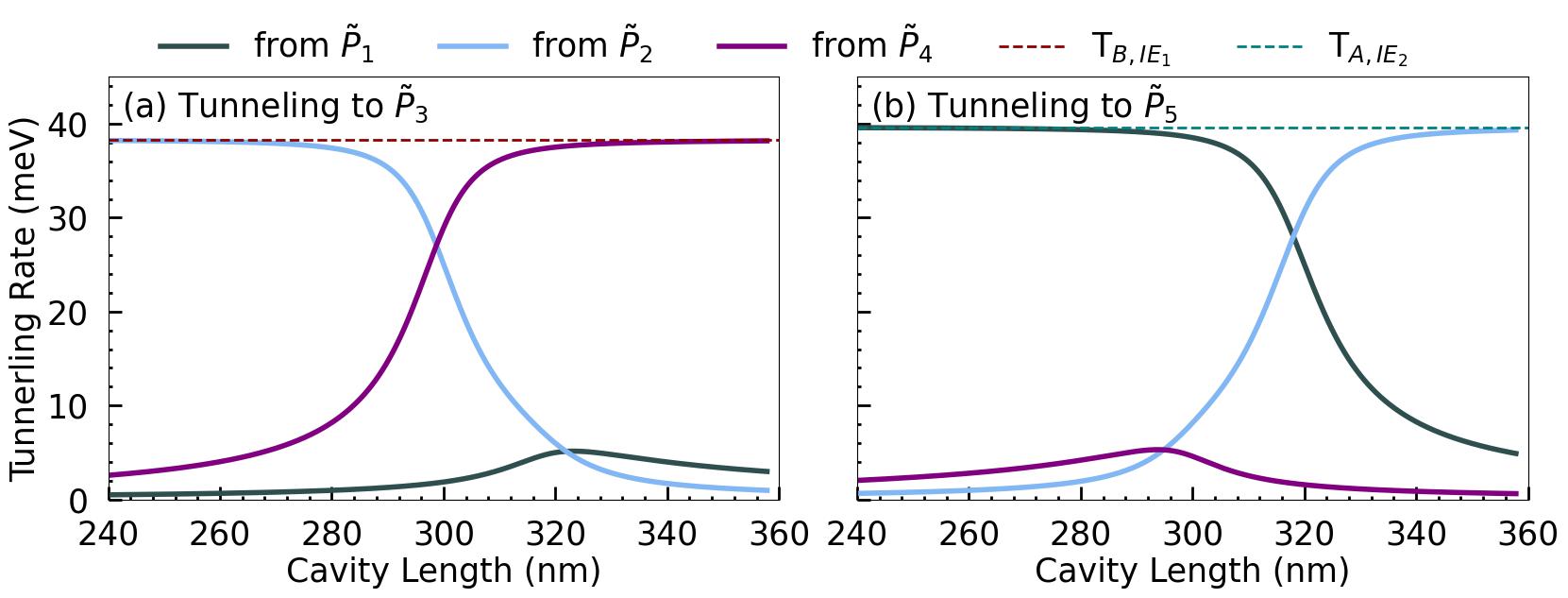}
    \caption{Tunneling between non-hybridized polaritons as a function of cavity length for the (a) $\tilde{P}_3$ and (b) $\tilde{P}_5$ polaritons. Furthermore, $T_\mathrm{B,IE_1}$ and $T_\mathrm{A,IE_2}$ (dashed horizontal lines) are shown and correspond to the  bare exciton tunneling matrix.}
    \label{fig:new_tunnel}
\end{figure}

In Fig. \ref{fig:new_tunnel}, a plot of the six relevant tunneling matrix elements is shown as a function of cavity length. The form of these resemble standard Hopfield coefficients (see Ref. \citenum{Fitzgerald2022}). If we look at the exemplary case of tunneling to polariton $\tilde P_3$ in Fig. \ref{fig:new_tunnel}(a), we see that in the limit of small cavities the tunneling from $\tilde P_2$ converges towards the bare tunneling between the IE$_1$ and B exciton (brown dashed line). This is expected because $\tilde P_2$ limits towards the B exciton for small cavity lengths (cf. Fig. 3(b) from the main text), while $\tilde P_3$ only consists of the IE$_1$ exciton. In the limit of long cavities, $\tilde P_2$ converges towards the A exciton (blue dashed line), which does not couple to the IE$_1$ exciton, meaning that the tunneling from $\tilde P_2$ to $\tilde P_3$ limits towards zero. Similar behavior can be seen for the tunneling from $\tilde P_4$ and $\tilde P_1$ to $\tilde P_3$, as well for tunneling from $\tilde P_1$, $\tilde P_2$ and $\tilde P_3$ to $\tilde P_5$ in Fig. \ref{fig:new_tunnel}(b).

\section{Two-exciton model}
\begin{figure}[t]
    \centering
    \includegraphics[width=1\textwidth]{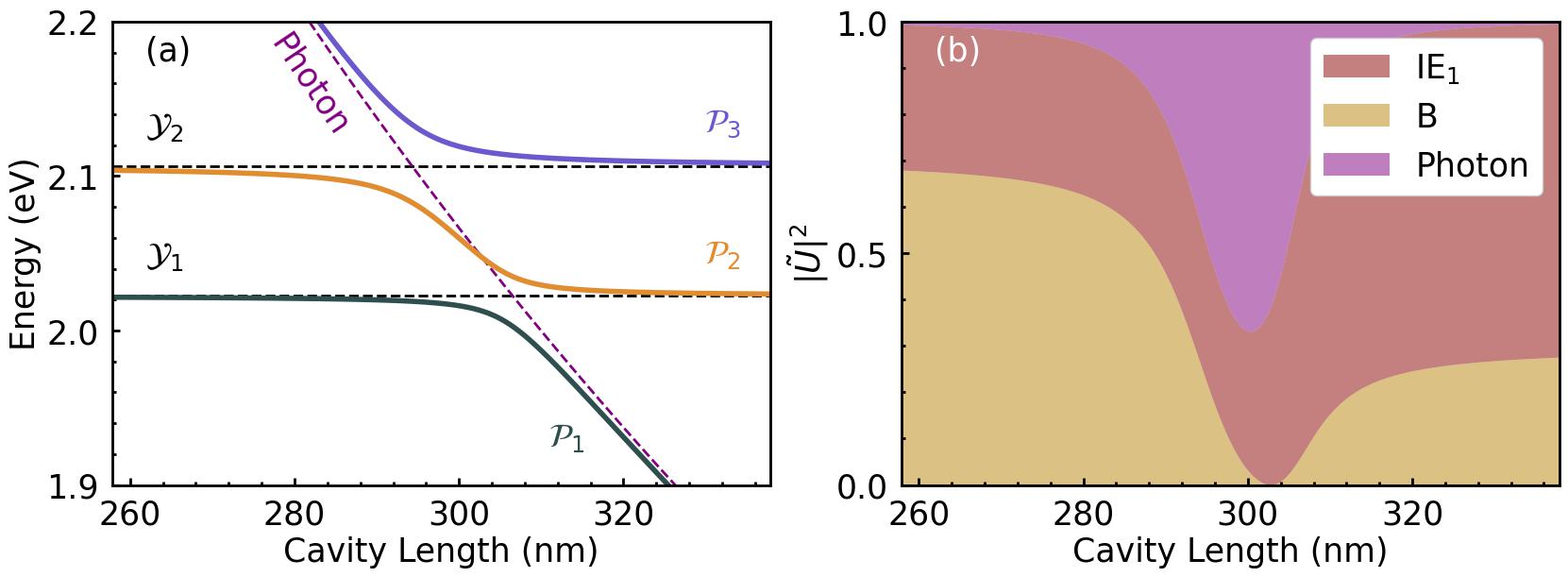}
    \caption{(a) Polariton dispersion for the two-exciton model, similar to Fig. 3(a) from the main text. (b) Generalized Hopfiled coefficents of the middle polariton branch $\mathcal{P}_2$ as a function of cavity length}
    \label{fig:2exc_model}
\end{figure}

Here, we use a simplified two-exciton-model, considering only the B and IE$_1$ excitons, to demonstrate that the photon-mediated dehybridization effect reported in the main text is a robust phenomenon. This simplification leaves the electronic hybridization unchanged since the matrix for this is already decoupled in two $2\times2$ matrices, giving us two hybrid excitons, corresponding to Y$_2$ and Y$_3$ from the main text, which we will call $\mathcal{Y}_1$ and $\mathcal{Y}_2$ here. In this fictional system, there are three polariton branches, which are denoted by $\mathcal{P}_n$. The energy of these polariton branches as a function of cavity length is shown in Fig. \ref{fig:2exc_model}(a).

We focus on the middle polariton branch, which corresponds to P$_3$ in the main text. As there is only one intralayer exciton, at a certain cavity length one can achieve perfect dehybridization, i.e.  the interlayer contribution to the total exciton component reaches 100\% (Fig.\ref{fig:2exc_model}(b)), i.e. we obtain a pure interlayer exciton polariton. 
To demonstrate this, we study  the generalized Hopfield coefficients. For a pure interlayer exciton polariton,  the  generalized intralayer Hopfield coefficients are expected to vanish. They read in the simplified two-exciton systems 
 \begin{align}
      \tilde{U}_\mathrm{B}^n&=C_\mathrm{B}^{\mathcal{Y}_1}U_{\mathcal{Y}_1}^n+C_\mathrm{B}^{\mathcal{Y}_2}U_{\mathcal{Y}_2}^n\;\text{.}
\label{genhop}
 \end{align}
For these coefficients to become zero, the following equation must be satisfied
$
    \frac{C_\mathrm{B}^{\mathcal{Y}_1}}{C_\mathrm{B}^{\mathcal{Y}_2}}=-\frac{U_{\mathcal{Y}_2}^n}{U_{\mathcal{Y}_1}^n}\;\text{.}$
This shows that if the mixing coefficients $C$ in the denominator and the numerator have the same sign, the normal Hopfield coefficients $U$ can not, and vice versa. The mixing coefficients are given by the well known result for the eigenvectors of a $2\times2$ matrix
\begin{subequations}
\begin{align}
    C_\mathrm{B}^{\mathcal{Y}_1}&=\mp\sqrt{\frac{E_{\mathcal{Y}_1}-E_\mathrm{IE_1}}{2E_{\mathcal{Y}_1}-E_\mathrm{IE_1}-E_\mathrm{B}}}\label{eq:hyb_ana1}\\
    C_\mathrm{B}^{\mathcal{Y}_2}&=\sqrt{\frac{E_{\mathcal{Y}_2}-E_\mathrm{IE_1}}{2E_{\mathcal{Y}_2}-E_\mathrm{IE_1}-E_\mathrm{B}}}
    \label{eq:hyb_ana2}
\end{align}
\end{subequations}
The sign of an eigenvector can be chosen freely while still keeping everything normalized, so it is only important that the two components of the eigenvector of the lower eigenvalue ($\mathcal{Y}_1$) are different, while for the higher eigenvalue ($\mathcal{Y}_2$) both components have the same sign. We note that this means only one of these mixing coefficients has an opposite sign to the remaining ones. The regular Hopfield coefficients in the reduced two-exciton system can be expressed as \cite{vasilevskiy2015exciton}
\begin{align}
    U_{\mathcal{Y}_1}^n&=-\frac{\Delta_{\mathcal{Y}_2}^n g^{(\mathcal{Y})}_{1}}{{\Delta_{\mathcal{Y}_1}^n}^2{\Delta_{\mathcal{Y}_2}^n}^2+{\Delta_{\mathcal{Y}_1}^n}^2\left(g^{(\mathcal{Y})}_{2}\right)^2+{\Delta_{\mathcal{Y}_2}^n}^2\left(g^{(\mathcal{Y})}_{1}\right)^2}\;\text{,}\\
     U_{\mathcal{Y}_2}^n&=-\frac{\Delta_{\mathcal{Y}_1}^n g^{(\mathcal{Y})}_{2}}{{\Delta_{\mathcal{Y}_1}^n}^2{\Delta_{\mathcal{Y}_2}^n}^2+{\Delta_{\mathcal{Y}_1}^n}^2\left(g^{(\mathcal{Y})}_{2}\right)^2+{\Delta_{\mathcal{Y}_2}^n}^2\left(g^{(\mathcal{Y})}_{1}\right)^2}\;\text{.}
\end{align}
with $\Delta_{\mathcal{Y}_i}^n$ as the energy difference between the polariton branch $n$ and the exciton $\mathcal{Y}_i$. Together with the definition of the light-exciton coupling strength in the hybrid basis (from the main text) $
    g_\eta^{(\mathcal{Y})}= C_\mathrm{B}^\eta g_\mathrm{B}^{(X)}+C_\mathrm{IE_1}^\eta g_\mathrm{IE_1}^{(X)}
$
and the vanishing coupling of the interlayer  excitons ($g_\mathrm{IE_1}^{(X)}=0$), the zero-condition for Eq.  (\ref{genhop}) can be rearranged to
\begin{align}
    \frac{\left(C_\mathrm{B}^{\mathcal{Y}_1}\right)^2}{\left(C_\mathrm{B}^{\mathcal{Y}_2}\right)^2}&=-\frac{E^{(\mathcal{Y})}_{1}-E^n}{E^{(\mathcal{Y})}_{2}-E^n}\label{eq:zero_cond2}\;\text{.}
\end{align}
For the middle polariton branch, either the denominator or the numerator on the right side of Eq. (\ref{eq:zero_cond2}) is negative and thus the entire expression  is always positive. For the left side we have two mixing coefficients squared, which is always positive, so the equality can be satisfied. Furthermore,  if the polariton branch is near $\mathcal{Y}_1$ the right side goes to zero and near $\mathcal{Y}_2$ to $+\infty$. The intermediate value theorem from calculus now tells us that there is always a certain cavity length for this equation to be true. As a result, the intralayer B component of the generalized Hopfield coefficient of the middle branch can always be found to vanish.
A very similar calculation can be done for the interlayer component of the generalised Hopfield coefficient. In this case one arrives at the equation
\begin{align}
    \frac{C_\mathrm{IE_1}^{\mathcal{Y}_1}C_\mathrm{B}^{\mathcal{Y}_1}}{C_\mathrm{IE_1}^{\mathcal{Y}_2}C_\mathrm{B}^{\mathcal{Y}_2}}&=-\frac{E^{(\mathcal{Y})}_{1}-E^n}{E^{(\mathcal{Y})}_{2}-E^n}\label{eq:zero_cond3}\;\text{.}
\end{align}
On the left side we have now all four mixing coefficients from Eqs. (\ref{eq:hyb_ana1}) and (\ref{eq:hyb_ana2}). As previously mentioned one of these has a different sign compared to the others, making the left side always negative. The right side is again always positive, which means Eq. (\ref{eq:zero_cond3}) can never be true. In summary, although it may look symmetric in the IE$_1$ and B components only the B exciton is responsible for the shared-out coupling, breaking this symmetry and leading to a maximum of 100\% in the excitonic interlayer component in this simplified model.

We have explored a range of different homobilayer configurations and found that this two-exciton model is an excellent approximation for R$_h^h$-stacked \ce{WSe2} homobilayer. This is because the different tunneling channels for R-type structures, and the much larger spin-orbit splitting of the valence band for W-based TMDs \cite{Kormanyos}, lead to two effectively decoupled two-exciton systems. Thus, for this material configuration we find two branches which both show a near $100\%$ interlayer maximum. The price paid though is a weaker interlayer hybridization and consequentially a weaker of Rabi splitting of about 17.7\,meV and 14.1\,meV at the energies of the most interlayer-like hybrid exciton polaritons.

\begin{figure}[t]
    \centering
    \includegraphics[width=1\textwidth]{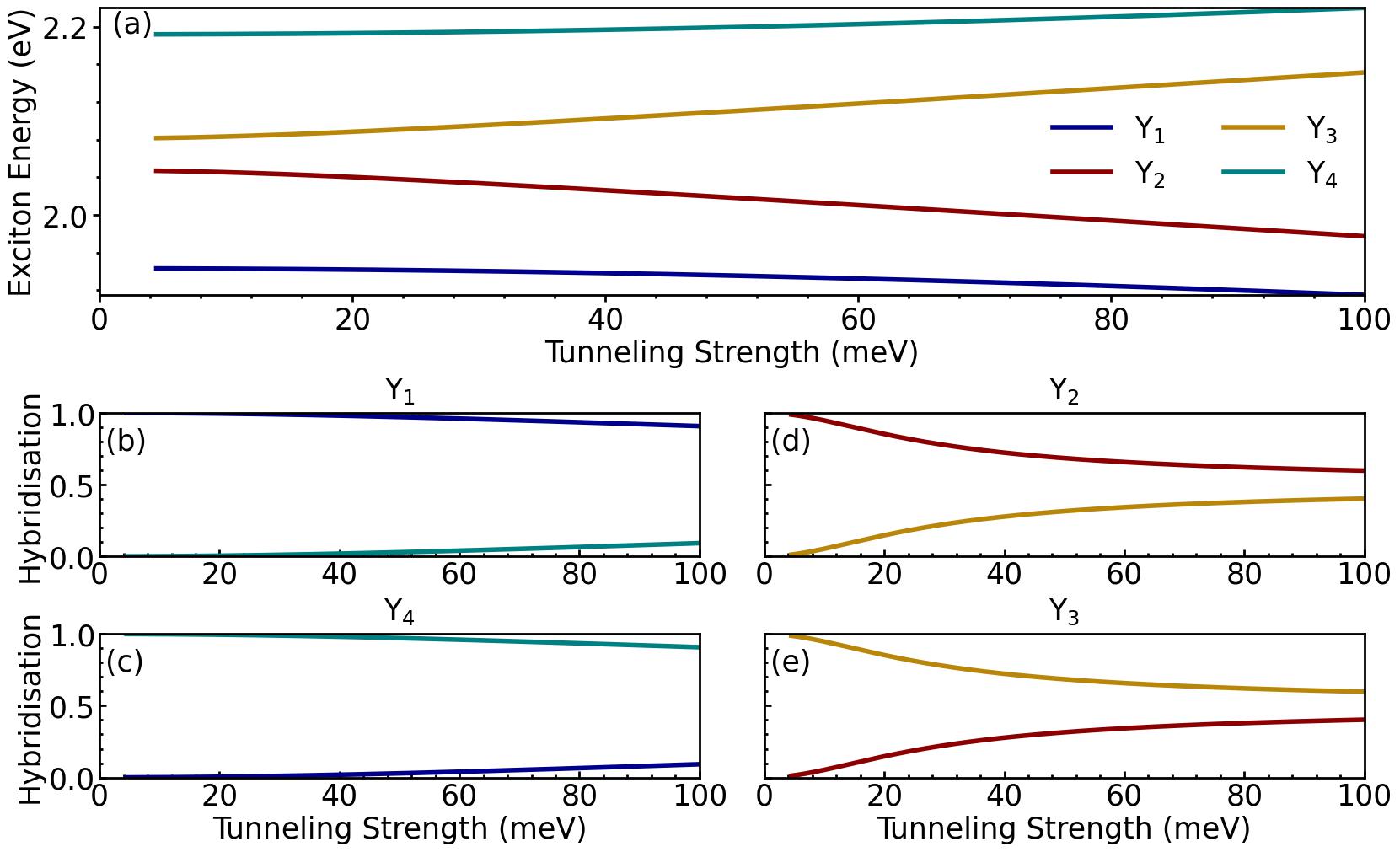}
    \caption{(a) Energy of hybridized excitons in H$_h^h$-stacked MoS$_2$ bilayers as a function of the hole tunneling strength. (b)-(e) Contributions from the different bare excitons to the hybridized excitons as a function of tunneling.  Note that the colors for bare excitons are the same as in fig. \ref{fig:All_hop}.}
    \label{fig:A_exc_tunnel}
\end{figure}

\section{Tunneling strength}
In this section we discuss the dependence on the tunneling strength. Experimental studies showing a reduced (enhanced) interlayer coupling by using pressure (tensile strain) \cite{zhu2022exchange} inspired us to vary the interlayer tunneling rate in the MoS$_2$ bilayer to gain more insights into the photon-induced exciton dehybridization effect. It is clear that a higher tunneling strength leads to a larger energy shift of the hybrid excitons relative to the bare excitons, as well as a larger degree of electronic hybridization. But a higher tunneling also changes the energy of the hybrid excitons relative to one another, which has an impact on the photon-induced mixing of hybrid excitons in the strong coupling regime. The shift of the hybrid excitons with tunneling strength is shown in Fig. \ref{fig:A_exc_tunnel}(a), demonstrating that Y$_1$ (Y$_4$) is energetically closer to Y$_2$ (Y$_3$) for lower tunneling strengths. One can also observe in Fig. \ref{fig:A_exc_tunnel}(b)-(e) that the degree of hybridization of Y$_2$ and Y$_3$ is much more affected by the tunneling strength compared to Y$_1$ and Y$_4$.

\begin{figure}[b!]
    \centering
    \includegraphics[width=0.75\textwidth]{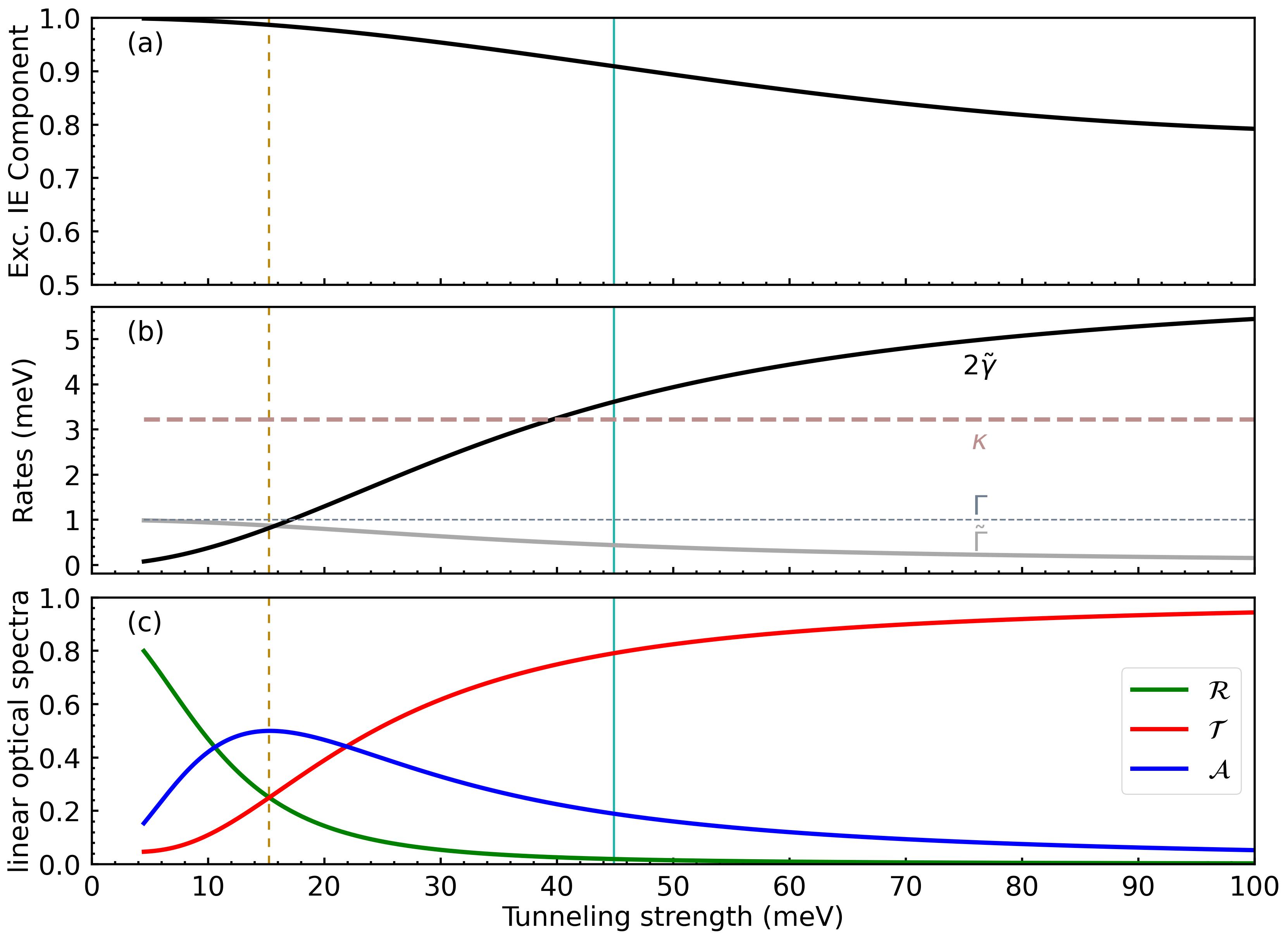}
    \caption{ \textbf{(a)} Interlayer component of the middle polariton branch P$_3$ as a function of tunneling strength for the cavity length at which dehybridization was observed in the main text. \textbf{(b)} Different decay rates as a function of tunneling strength. \textbf{(c)} Reflection, transmission and absorption as a function of tunneling. The vertical yellow dashed lines show the maximum in absorption, while the vertical turquiose line shows the actual tunneling strength of the hole of 44.9\,meV.}
    \label{fig:tunnel_sweep}
\end{figure}

In Fig. \ref{fig:tunnel_sweep}(a) we can observe how the interlayer excitonic component of the middle polariton branch P$_3$ (Fig. 3(a) from the main text) decreases for higher tunneling strengths. To explain this, we should keep in mind that the photon-mediated dehybridization cancels out only the intralayer B component, while the remaining intralayer A component hinders the complete cancellation. Now, with a lower tunneling strength the most A-like hybrid exciton, Y$_1$, is further away from the middle polariton branch P$_3$ between Y$_2$ and Y$_3$, as shown in Fig. \ref{fig:A_exc_tunnel}(a). This, together with a reduced coupling due to a reduced degree of hybridization of Y$_2$ (Fig. \ref{fig:A_exc_tunnel}(d)) leads to a decreased contribution of the A exciton to P$_3$ for lower tunneling strengths, making a near-100\% interlayer dipolariton possible. In Figs. \ref{fig:tunnel_sweep}(b) and (c) we show the polaritonic decay channels and linear optical spectra as a function of cavity length, which are disconnected from the dehybridization. For the loss parameters we chose in this work, the absorption peaks at $0.5$ for lower tunneling strengths. By tuning the loss parameters, e.g., changing the exciton decay rate with temperature, the maximum absorption coincides with the actual tunneling strength of 44.9\,meV at the point of dehybridization.

\section{Derivation of the polariton-polariton interaction}
In order to gain microscopic access to polariton-polariton interactions, considering interactions between exciton-polaritons which exhibit a large interlayer exciton component, we start from a bosonic exciton-exciton Hamiltonian \cite{erkensten2021exciton}:
\begin{equation}
    \hat{H}_{x-x}=\frac{1}{2}\sum_{L, L', \bm{q}}D^{LL'L'L}_{\bm{q}} \hat{X}^{\dagger}_{L, \bm{Q+q}}\hat{X}^{\dagger}_{L', \bm{Q}'-\bm{q}}\hat{X}_{L', \bm{Q}'}\hat{X}_{L, \bm{Q}} \ ,
    \label{hamilton}
\end{equation}
where $L, L'$ are layer indices, $\bm{Q}, \bm{Q}', \bm{q}$ are momenta and $D_{\bm{q}}^{LL'L'L}$ is the interlayer exciton-exciton interaction matrix element. Note that we 
focus on the dominant interaction for interlayer excitons, that is the dipole-dipole repulsion \cite{kyriienko2012spin, erkensten2022microscopic}. Exchange interactions involving the individual exchange of fermionic constituents between excitons only yield a small quantitative correction to the interaction strength for interlayer excitons. In the long wavelength limit (corresponding to small momenta), the interaction between identical interlayer excitons reads \cite{erkensten2022microscopic}
\begin{equation}
    D^{\mathrm{IE-IE}}=\frac{de^2}{\epsilon_0 \epsilon_{\perp}} \ , 
\end{equation}
where $d$ is the TMD layer separation (here taken as the TMD layer thickness, $d\approx{0.65}$ nm), $e$ is the electric charge and $\epsilon_{\perp}$ is the out-of-plane component of the dielectric tensor of the TMD. 

The Hamiltonian in Eq. \eqref{hamilton} is now expanded into the hybrid exciton basis using the basis transformation $\hat{X}_{L, \bm{Q}}=\sum_\eta Y_{\eta, \bm{Q}} C_{L, \bm{Q}}^{\eta*} $ \cite{brem2020hybridized}, resulting in a hybrid exciton-exciton Hamiltonian containing the hybrid exciton-exciton interaction 
\begin{equation}
D_{\bm{Q}, \bm{Q}', \bm{q}}^{\eta_1\eta_2\eta_3\eta_4}=\sum_{LL'}D_{\bm{q}}^{LL'L'L}C_{L,\bm{Q+q}}^{\eta_1}C_{L',\bm{Q}'-\bm{q}}^{\eta_2}C^{*\eta_3}_{L',\bm{Q}'}C^{*\eta_4}_{L,\bm{Q}} \ . 
\end{equation}
Now, as a final step, we transform the hybrid exciton-exciton Hamiltonian to the (hybrid) exciton polariton basis \cite{Fitzgerald2022}, enabling us to get an effective dipolariton-dipolariton Hamiltonian
\begin{equation}
     \hat{H}_{p-p}=\sum_{\substack{n_1\cdots n_4\\\bm{Q},\bm{Q}',\bm{q}}}\tilde{D}_{\bm{Q}, \bm{Q}', \bm{q}}^{n_1n_2n_3n_4} \hat{P}_{n_1,\bm{Q}+\bm{q}}^{\dagger} \hat{P}_{n_2,\bm{Q}'-\bm{q}}^{\dagger} \hat{P}_{n_3,\bm{Q}'}\hat{P}_{n_4,\bm{Q}}\ , 
\end{equation}
expressed in terms of the polariton operators $P_{n, \bm{Q}}=\sum_{\eta} Y_{\eta, \bm{Q}} U^n_{\eta}$, $U^{\eta}_n$ being Hopfield coefficients with $n$ enumerating the polariton branch. We have now obtained the dipolariton-dipolariton interaction strength 
\begin{equation}
       \tilde{D}_{\bm{Q}, \bm{Q}', \bm{q}}^{n_1n_2n_3n_4}=\sum_{LL'}D_{\bm{q}}^{LL'L'L}\tilde U_{L,\bm{Q}+\bm{q}}^{n_1}\tilde U_{L',\bm{Q}'-\bm{q}}^{n_2}\tilde U_{L',\bm{Q}'}^{n_3*}\tilde U_{L,\bm{Q}}^{n_4*}\ , 
       \label{dpdp}
\end{equation}
where we absorbed the excitonic mixing coefficients in the \emph{generalized} Hopfield coefficients $\tilde U_{L, \bm{Q}}^{n(*)}=\sum_\eta C_{L, \bm{Q}}^{\eta(*)}U_\eta^{n(*)}$. As we only focus at normal incidence, i.e. zero momentum of excitons ($\bm{Q}= \bm{Q}'= \bm{q}=0$), Eq. \eqref{dpdp} can be simplified, while only considering interactions between polaritons from the same branch ($n_1=n_2=n_3=n_4=n$), to 
\begin{equation}
    D^n=\sum_{L,L'= \mathrm{IE}_1, \mathrm{IE}_2}D^{LL'L'L}|\tilde U_{L}^n|^2|\tilde U_{L'}^n|^2 \ , 
\end{equation}
defining $D^n\equiv D^{n}_{\bm{0}, \bm{0}, \bm{0}}$. As shown in Fig. \ref{fig:All_hop}(b), the considered polariton branches (P$_1$...P$_5$) predominantly have contributions only from a single interlayer exciton species, IE1 or IE2, independent of cavity length. This implies that the mixed terms ($L\neq L'$) in the expression above can be ignored, yielding the final form of the polariton-polariton interaction
 \begin{equation}
    D^n=D^{\mathrm{IE-IE}}\left(|\tilde U_{\mathrm{IE_1}}^n|^4+|\tilde U_{\mathrm{IE_2}}^n|^4\right)\ .
\end{equation}
This expression has already been discussed in the main text.

\bibliography{ref}